\documentclass[]{aa}  
\usepackage{graphicx}
\usepackage{txfonts}
\usepackage{hyperref}
\usepackage{subcaption}
\usepackage{natbib}

\newcommand{\rvir}{r_{\rm vir}}

\newcommand{\orcidd}[1]{\href{https://orcid.org/#1}{\protect\includegraphics[width=8pt]{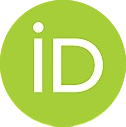}}}

\begin{document}

   \title{The long-term evolution of ultra-faint dwarf galaxies and observational implications}
   \titlerunning{UFD long-term evolution}
\authorrunning{Flammini Dotti et al}

   \author{Francesco Flammini Dotti\fnmsep\thanks{mailto:ff2415@nyu.edu}
          \inst{1,2,3,4}\orcidd{0000-0002-8881-3078}, 
          Roberto Capuzzo-Dolcetta\inst{1,5,6,7}\orcidd{0000-0002-6871-9519},
          Giovanni Carraro\inst{8}\orcidd{0000-0000-0000-0000},
          Alessandro Alberto Trani\inst{9,10}\orcidd{0000-0001-5371-3432} 
          \and
          Rainer Spurzem\inst{4, 10,11}\orcidd{0000-0003-2264-7203}
          }

   \institute{
                Dipartimento di Fisica, Sapienza, Universit\'a di Roma, P.le A.Moro 2, I-00185, Roma, Italy
                \and Department of Physics, New York University Abu Dhabi, PO Box 129188 Abu Dhabi, UAE
\and Center for Astrophysics and Space Science (CASS), New York University Abu Dhabi, PO Box 129188, Abu Dhabi, UAE
\and Astronomisches Rechen-Institut, Zentrum f\"ur Astronomie, University of Heidelberg, M\"onchhofstrasse 12--14, 69120, Heidelberg, Germany
             \and INFN, Sezione di Roma, Sapienza, Universit\'a di Roma, P.le A.Moro 2, I-00185, Roma, Italy
             \and INAF, Observatory of Roma at Monteporzio Catone, via Frascati, 33, I-00078, Monteporzio Catone, Italy 
             \and
             Centro Ricerche Enrico Fermi,
             Piazza del Viminale 1,
             I-00184 Roma, Italy
             \and
             Dipartimento di Fisica e Astronomia, Universit\'a di Padova, Vicolo Osservatorio 3, I-35122 Padova, Italy
             \and
                 Departamento de Astronom\'ia, Facultad Ciencias F\'isicas y Matem\'aticas, Universidad de Concepci\'on, Avenida Esteban Iturra, Casilla 160-C, Concepci\'on, 4030000, Chile
\and
National Institute for Nuclear Physics – INFN, Sezione di Trieste, I-34127, Trieste, Italy
\and
  National Astronomical Observatories, Chinese Academy of Sciences, 20A Datun Rd., Chaoyang District, 100101, Beijing, China
   \and
             Kavli Institute for Astronomy and Astrophysics, Peking University, Yiheyuan Lu 5, Haidian Qu, 100871, Beijing, China}

   \date{Received September 15, 1996; accepted March 16, 1997}

  \abstract 
  {In the Local Group, dwarf spheroidal galaxies and ultra-faint dwarf galaxies (UFDs) exhibit large velocity dispersions. These values are generally attributed to the presence of substantial amounts of dark matter (DM), in line with predictions of the standard model of galaxy formation. However, alternative explanations exist, such as non-virialized dynamical states induced by tidal interactions, the presence of stellar streams, and the artificial inflation of the velocity dispersion caused by binary-star orbital motion.}
{We studied the dynamical evolution of UFDs using purely stellar (``dry'') dynamics, without invoking DM. We dynamically evolved our systems up to a Hubble time and compared our results with observational studies and previous theoretical work.}
{We employed direct high-precision N-body simulations performed with the \texttt{NBODY6++GPU} code. We explored the role of binaries in inflating the velocity dispersion of low-mass host galaxies. We also present both the stellar and dynamical evolution of the stellar population, which is necessary to properly interpret our results.}
{We find that, in all our models, the UFD remains globally quasi-stationary for approximately 3000 Myr. Subsequently, the system undergoes mass segregation and experiences a phase resembling core collapse. Red giants and white dwarfs are found to play significant, but distinct, roles. Red giants provide the dominant contribution to the luminosity, whereas white dwarfs constitute the largest fraction of the nonluminous component, accounting for approximately 13\% of the total stellar population. Finally, if not properly taken into account, velocity dispersion measurements can be strongly biased by the presence of a significant binary population, which can lead to substantial overestimates of velocity dispersion in UFDs.}


\maketitle

\section{Introduction}

Our Galaxy hosts a significant number of dwarf galaxies with diverse properties. These systems are commonly classified as dwarf spheroidal, dwarf elliptical, and dwarf irregular galaxies. The least luminous members of this population constitute the class of ultra-faint dwarf galaxies (UFDs). UFDs are defined as dwarf galaxies with an integrated V-band magnitude ($M_V$) $> -7.7 $  ($L \lesssim 10^5\,L_\odot$) \citep{sim19}. The number of known UFDs has increased significantly since the Sloan Digital Sky Survey (SDSS) began identifying such objects \citep{wil05a,wil05b}. Over the subsequent two decades, the SDSS, the Dark Energy Survey (DES), and the Panoramic Survey Telescope and Rapid Response System (Pan-STARRS) have discovered approximately eight times more dwarf satellites of the Milky Way (MW) than were known prior to the operation of these surveys. To date, the number of MW dwarf galaxies has exceeded 80 \citep{sim19}. As of today, few extragalactic sources with properties similar to those of UFDs have been discovered. We report the discovery of PegasusW \citep{Sand2022_PegasusW} in the Local Group, Leo M and Leo K \citep{Battaglia2023_LeoMK} just outside of the MW halo, and six others orbiting NGC 253 \citep{Sand2021_NGC253} and NGC300 \citep{MutluPakdil2022_NGC300}, three for each galaxy. Note that these were detections just above the sky background limits and need confirmation (especially the megaparsec-distance source).

In the review by \citet{sim19}, it is implied that UFDs are comparable in luminosity (and stellar mass) to Galactic globular clusters but that their half-light radii are at least an order of magnitude larger. This definition is consistent with that adopted by \citet{bul17}. The main characteristics of UFDs, as summarized by \citet{sim19}, are as follows: (a) the stellar kinematics of UFDs indicate the presence of large dark matter (DM) fractions; (b) all dwarf galaxies, with the exception of the lowest-luminosity UFDs, are larger than any known globular cluster; (c) within individual UFDs, the abundances of Fe and $\alpha$-elements exhibit substantial internal spreads, indicative of extended star formation and self-enrichment; (d) UFDs follow a luminosity--metallicity relation \citep{hidalgo2017}, whereas globular clusters do not; and (e) the abundance patterns of certain elements in UFDs resemble those observed in brighter dwarf galaxies and differ from the light-element abundance correlations characteristic of globular clusters.

In this work we focus primarily on point (a). Velocity dispersion measurements are available for a limited number of MW UFDs (26 of the 42 candidates reported in \citealt{sim19}), as well as for Pegasus~IV \citep[reported in][]{cer23}, yielding a total sample of 27 systems. We adopted the updated measurements of \citet{chi22} for Grus~I. The resulting mean velocity dispersion is $\langle \sigma \rangle = 4.66\,\mathrm{km\,s^{-1}}$. In \citet{sim19}, the dynamical mass enclosed within the three-dimensional half-light radius ($R_{1/2}$), denoted $M_{1/2}$, was computed following \citet{wol10} as
\begin{equation}
M_{1/2} = 930\,\sigma_{1\mathrm{D}}^2\,R_{1/2},
\end{equation}
where $\sigma_{1\mathrm{D}}$ is the one-dimensional velocity dispersion in $\mathrm{km\,s^{-1}}$ and $R_{1/2}$ is the projected two-dimensional half-light radius. The derived mean mass-to-light ratio for the UFD sample is $\langle M_{1/2}/L_V \rangle = 459.95\,M_\odot/L_\odot$, an extremely high value, approximately an order of magnitude higher than that of classical dwarf galaxies ($M_V < -7.7$), which have an average $M/L \approx 48.83\,M_\odot/L_\odot$. A similar trend is observed for M31 satellites, with $\langle M/L \rangle = 449.4\,M_\odot/L_\odot$ for UFDs and $\langle M/L \rangle = 55.34\,M_\odot/L_\odot$ for brighter dwarfs.

These results mirror earlier findings for Local Group dwarf spheroidal galaxies (dSphs; \citealt{2011MNRAS.411.2118A}). These systems, typically located at Galactocentric distances exceeding 70~kpc and composed of old, metal-poor stellar populations, are therefore structurally similar to Galactic globular clusters. They exhibit no clear evidence of global rotation and are thus not rotationally supported, implying that mass estimates must rely on spectroscopic measurements of their velocity dispersions \citep{1993AJ....105..510M}.

Several studies of classical dSphs in the MW halo (e.g., Fornax, Sculptor, Ursa Minor, Draco, Leo~I, Leo~II, Sagittarius, Sextans, and LGS~3) have shown that the observed velocity dispersion, $\sigma_{\mathrm{obs}}$, is significantly higher than the value expected if these systems were simple scaled-up versions of globular clusters, for which $\sigma \sim 1$--$3\,\mathrm{km\,s^{-1}}$  (\citeauthor{1997ASPC..116..259M}, \citeyear{1997ASPC..116..259M}). Subsequent work by \cite{2007ApJ...670..313S} confirmed observed dispersions in the range 3.3--$7.6\,\mathrm{km\,s^{-1}}$, challenging earlier claims of an upper limit $\sigma_{\mathrm{obs}} \leq 7\,\mathrm{km\,s^{-1}}$  (\citealt{2008IAUS..244...44W}) and motivating revisions of the inferred mass scale of these systems.

Several scenarios have been proposed to explain the unusually large velocity dispersions observed in dSphs and UFDs. Early studies (e.g., \citealt{1988IAUS..130..409A}) invoked the presence of substantial DM halos, consistent with predictions of the standard cosmological model. Alternative interpretations suggested that dSphs might be out of virial equilibrium due to ongoing tidal disruption. However, the importance of Galactic tides has been questioned based on the observed luminosity--metallicity relation \citep{Kirby2008} and the lack of clear observational signatures of tidal features in many systems. For example,  \citeauthor{2009ApJ...692.1464G} (\citeyear{2009ApJ...692.1464G}) found no evidence of tidal tails or induced rotation in Segue~I, rejecting the hypothesis that it is a disrupted globular cluster formerly associated with the Sagittarius stream (\citeauthor{2007ApJ...654..897B}, \citeyear{2007ApJ...654..897B}).

Another possibility is that the elevated velocity dispersions arise from binary-star orbital motion. Although the tidal disruption scenario has largely been disfavored, the role of binaries remains an active topic of investigation. \cite{1997ASPC..116..259M} argued that unresolved binaries are unlikely to fully account for the inflated dispersions observed in classical dSphs, which are therefore still considered DM-dominated. However, their impact may be non-negligible in UFDs, the low-luminosity counterparts of classical dSphs  \citep{2010ApJ...722L.209M,2018AJ....156..257S}. Despite recent expansions of the UFD sample \citep{2018A&A...620A.155M}, small-number statistics and the lack of extensive multi-epoch spectroscopic observations continue to hinder robust constraints on binary fractions and orbital parameter distributions \citep{2010ApJ...722L.209M,2018AJ....156..257S}.
In this context, it is worth mentioning the case of Segue II, for which the inflation of the velocity dispersion has been extensively debated \citep{2009MNRAS.397.1748B,2013ApJ...770...16K}. Unfortunately, only in a limited number of cases do the available spectroscopic data allow meaningful constraints to be placed on the binary fraction (e.g., the UFD galaxy Reticulum II; \citealt{2019MNRAS.487.2961M}). Other recent studies also suggest the importance of wide binaries in UFDs \citep{ElBadry2019_WideBinaries}, and a resolved stellar population will allow us to place constraints on the binary fractions and eventually the DM abundance \citep{MutluPakdil2023_UFDBinaries}. For this reason, modeling approaches based on Monte Carlo simulations and Bayesian analyses have been widely adopted. To date, most models have aimed to reproduce the observed velocity dispersion of classical dSphs by varying both the binary fraction and the binary orbital parameters, and subsequently comparing the results with spectroscopic data to estimate the contribution of binaries to the observed velocity dispersion, $\sigma_{\mathrm{obs}}$, in UFDs \citep{2017AJ....153..254S,2018A&A...620A.155M}. Nevertheless, assumptions regarding the distributions of orbital parameters -- particularly orbital periods and semimajor axes -- represent a significant limitation in this framework. This motivates the development of theoretical models capable of making inferences about the binary populations in these systems that are as reliable and general as possible.
Assuming that both dSphs and UFDs are DM-dominated systems, which is currently the most widely supported scenario, we present in this paper an N-body study aimed at exploring the effects of different choices of binary orbital parameters using different binary fractions on the observed velocity dispersion of such galaxies.

The ultimate purpose of this work is to investigate the impact of binaries on the determination of the dynamical mass in the faintest MW satellites, with particular reference to \cite{2020ApJ...896..152R} regarding the methodology used to compute velocity dispersions. Finally, recent studies \citep{shirazi2025} suggest that baryonic explanations may be possible in some cases, although additional components (e.g., a central black hole) may be required to match the inferred dynamical mass in specific systems.

This paper is organized as follows. In Sect.~\ref{methodology} we describe the methodology and the simulation set-up. In Sect.~\ref{results} we present our results, and in Sect.~ \ref{conclusions} we summarize our main conclusions.

\section{Methodology and initial conditions}
\label{methodology}
Of the dwarf galaxies known in the local Universe, UFDs are the faintest. They are satellites of the MW that populate the lowest-luminosity tail of the galaxy luminosity function ($L \lesssim 10^5\,L_\odot$) and are commonly considered to be the oldest, most metal-poor, least chemically evolved, and most DM-dominated stellar systems known.

The distinction between dSphs and UFDs is somewhat arbitrary. Here, we adopted the classification proposed by \citet{bul17} and subsequently used by \citet{sim19}, defining UFDs as systems with integrated magnitudes $M_V \geq -7.7$ ($L \lesssim 10^5\,L_\odot$).

In this paper we present a set of direct N-body simulations of a prototype UFD composed of both single stars and binaries. The dynamical evolution of each model was followed with high precision using the \texttt{NBODY6++GPU} code, from time zero (after gas dissipation) up to one Hubble time.

The main goals of this paper are: (i) to investigate the long-term evolution of a UFD-like system in the presence of low, intermediate, and high binary fractions, thereby testing the assumption that such systems are effectively collisionless; and (ii) to quantify how binaries affect both the internal dynamics of the galaxy and the observational determination of its virial mass. This latter aspect extends previous studies \citep[e.g.,][]{ElBadry2018,2020ApJ...896..152R,2022ApJ...939....3P}, in which the impact of binaries on mass estimates was explored using statistical or semi-analytical approaches.

It is well established that the binary fraction increases toward lower metallicities \citep{moe19}. Several binaries have been identified in UFDs \citep[e.g.,][]{spe18}, and observational evidence suggests that the binary fraction may range from a few percent up to approximately 70\%. Combined with the growing evidence that binaries can significantly inflate the measured velocity dispersion in faint galaxies, this motivates a detailed investigation of their dynamical role, especially given the claimed extreme DM content of these systems \citep{irwin1995}.

Binary-induced velocity dispersion inflation has been observed, for example, in Very Large Telescope studies of Bo\"otes~I, Leo~IV, and Leo~V \citep{jen21}. In Bo\"otes~I and Leo~IV, the one-dimensional velocity dispersion decreases by approximately 0.2--$0.3\,\mathrm{km\,s^{-1}}$ once the spurious binary contribution is removed. However, we believe that due to the small-number statistics, observational data alone remain insufficient to robustly quantify the binary contribution to $\sigma$, which enters quadratically into virial mass estimates.

For this reason, numerical approaches have been adopted to assess the influence of binaries on velocity dispersion measurements. \citet{2020ApJ...896..152R} conducted a full, high-precision N-body study of low-mass stellar systems comparable to open clusters, while \citet{2022ApJ...939....3P} employed a statistical framework to study systems with sizes comparable to dSphs and UFDs. These authors found that, particularly in the faintest galaxies, binary-induced inflation of $\sigma$ can be substantial, potentially leading to virial mass overestimates of a factor of two or more for binary fractions of $\sim 40\%$ when multi-epoch observations are unavailable. Actually, multi-epoch are extremely rare and limited for UFDs \citep[e.g.,][]{Kirby2015_UFDMultiEpoch,Spencer2017_UFDBinaries,Spencer2018_ReticulumII,Jenkins2021_UFDSurvey,Buttry2022_TriangulumII, Simon2011_Segue1,Simon2015_RetII_Discovery}.

The full dynamical evolution presented in this paper should shed light on the debated question of the selective effect of the binary content of a UFD on the overall dynamics. Although UFDs are commonly treated as collisionless systems, the presence of a significant binary population complicates this picture. Interactions among different stellar components can alter the binary population, preferentially disrupting wide binaries while leaving hard binaries intact. This evolution has direct consequences for mass estimates derived from velocity dispersion measurements. For the evolution of wide binaries in dwarf galaxies, we refer to \citet{pen16} and \citet{liv23}.

Our data analysis follows the methodology described in \citet{2022ApJ...939....3P} and includes several definitions of the velocity dispersion: (1) a global dispersion, $\sigma_g$, computed by treating all stars as if they were single; (2) a dispersion $\sigma_{g,\mathrm{cm}}$, in which binaries contribute only through their center-of-mass velocities; and (3) a luminosity-weighted dispersion, $\sigma_L$, in which single stars are treated normally and binaries are weighted according to
\begin{equation}
v = \frac{L_A v_A + L_B v_B}{L_A + L_B},
\end{equation}
where $L_A$, $v_A$, and $L_B$, $v_B$ are the luminosities and speeds of the binary components \citep[see][]{2020ApJ...896..152R}.

The uncertainty in our work arises from both systematic and statistical sources. Systematic uncertainties include numerical integration errors as well as uncertainties in the photometric output derived from the mass--luminosity relations implemented in the code. For a detailed discussion of these effects, we refer to \citet{Kamlahetal2022a}.

\subsection{UFD initial conditions}

\begin{table}
        \caption{UFD properties.}
    \label{tab:ics}     
        \resizebox{0.5\textwidth}{!}{
    \begin{tabular}{|c|} 
    \hline
    Ultra-faint dwarf galaxy properties\\
                \hline
        \hline
        $N= 82 000$ stars      \\
         \hline
        $Q = 0.5$\\
         \hline
                    $\rvir$ = 50 pc \\
             \hline
            Plummer model density, Kroupa 2001 initial mass function \\ 
             \hline
            No external tidal field \\
             \hline
            Simulation time = 13\,700 Myr \\
             \hline
                        $t_{\rm rel, c} \approx 6\, 000 \rm\,Myr \ ; t_{\rm rel,hm} \approx 43\,  000\rm\, Myr$\\
             \hline
            $t_{\rm df, c} \approx 2\,000 \rm\,Myr \ ; t_{\rm df,hm} \approx 14\,200 \rm\, Myr$\\
             \hline
            Z=0.01\\
             \hline
                    $f_{b}$ = 0,\,0.1,\,0.2,\,0.3,\,0.4,\,0.5  of stars \\
                    Random coupling for binaries (1),(2) \\
                     \hline
                        Log uniform distribution of semimajor axis $a$ in the $0.1-50$ AU range        \\
             \hline
            Thermal distribution of eccentricities $e$\\
             \hline
             Model names: U0, U10, U20, U30, U40, U50, according to the binary fraction \\
                        \hline\hline
        \end{tabular}}
    \tablefoot{Settings of the UFD models and its binaries, illustrating the general properties and the binaries properties. $N$ is the number of stars, $Q$ is the virial ratio of the cluster, $\rvir$ is the virial radius, the Plummer model is the density profile distribution, $t_{\rm rel, c}$ and $t_{\rm rel, hm}$ are the relaxation time at the core radius (the radius where the density of the cluster is half of the central density) and the half-mass radius region (the numerical equivalent of the half-light radius, counting the number of stars instead of the luminosity).  $t_{\rm df, c}$ and $t_{\rm df, hm}$ are the dynamical friction timescales, which we assumed to be one-third of the relaxation time (see text). Z is the absolute metallicity of the UFD. \\
    For the binaries, we define $f_b$ as the fraction of the total number of stars that are in binary systems, so 50 \%, for $N=82\,000$ stars, means that 41\,000 stars are in a $20\,500$ binary systems.}
    \tablebib{
(1)~\citet{weidner2009}; (2) \citet{Wangetal2015}}
\end{table}

Table \ref{tab:ics} summarizes the settings of our UFD models, which are largely inspired by \cite{2022ApJ...939....3P}.
We sampled a Plummer model \citep{Plummer1911} with 82\,000 stars, adopting different binary fractions $f_b = N_b/N$, where $N_b$ is the number of binary systems and $N$ is the total number of stars. We explored values of $f_b$ from 0 to 0.5, in steps of 0.1.

The system was initially contained within a virial radius of $r_{\mathrm{vir}} = 50$~pc. Stellar masses were drawn from a \citet{kroupa} initial mass function over the range 0.08--$150\,M_\odot$, yielding a total initial stellar mass of $6.24 \times 10^4\,M_\odot$. The model was initialized in virial equilibrium and evolved dynamically for a time comparable to one Hubble time (13.7~Gyr).

Ultra-faint dwarf galaxies are generally regarded as collisionless stellar systems due to their low densities and their characteristic radii, which can be either  small or relatively large. This, despite the low velocity dispersion, implies very long two–body relaxation times.
The local relaxation time can be estimated as \citep[][Eq. 7.106]{binney2008}
\begin{equation}
t_{\mathrm{rel}} = \frac{0.34\,\sigma^3}{G^2 m_\ast \rho \ln \Lambda},
\end{equation}
where $\sigma$ and $\rho$ are the local one-dimensional velocity dispersion and mass density, $m_\ast$ is the mean stellar mass, and $\ln \Lambda$ is the Coulomb logarithm.
The Coulomb logarithm represents the cumulative effect of numerous weak gravitational encounters between stars. Formally, it arises from integrating over the range of impact parameters, $b$, that contribute to the scattering between two bodies, and is defined as $\ln \Lambda = \ln (b_{\rm max}/b_{\rm min})$. The lower cutoff, $b_{\rm min}$, is typically taken to be on the order of the impact parameter corresponding to a $90^\circ$ deflection, $b_{90} = 2G m_*/\sigma^2$, while the upper cutoff, $b_{\rm max}$, is set by the characteristic size of the system, usually the local scale length of the density distribution or the half-mass radius \citep[e.g.,][]{spitzer1987, binney2008}. Because the exact values of these cutoffs are somewhat ambiguous and system-dependent, $\ln \Lambda$ is often approximated as a constant in the range 5–15, depending on the stellar system under consideration. For globular clusters, $\ln \Lambda \simeq 10$ is commonly adopted, as it yields accurate order-of-magnitude estimates of the relaxation time.
Thus, following standard practice, we adopted $\ln \Lambda = 10$.

Under this assumption, we estimate a half-mass relaxation time of $t_{\mathrm{rel,hm}} \approx 43\,000$~Myr and a core relaxation time of $t_{\mathrm{rel,c}} \approx 6\,000$~Myr. Here, the core radius is defined as the three-dimensional radius at which the stellar density drops to half its central value. Although the relaxation time exceeds a Hubble time in most regions, the innermost regions experience relaxation on timescales shorter than the simulation duration. Consequently, we expect mild but non-negligible collisional effects to develop over the course of the evolution.

\begin{figure}
        \includegraphics[width=\columnwidth]{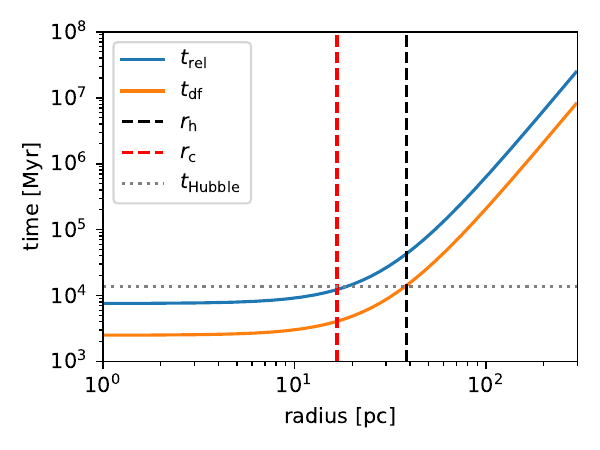}
        \caption{Initial relaxation (blue) and dynamical friction (orange) times of our UFD Plummer model as functions of radius. The vertical dashed red and black  lines denote the initial core and half-mass radius of the adopted Plummer model. For reference, the horizontal gray line marks the Hubble time.}\label{fig:trelax}
\end{figure}

Binary systems segregate on the dynamical friction timescale \citep[e.g.,][]{binney2008},
\begin{equation}
t_{\mathrm{df}} = 0.66\,\frac{m_\ast}{m_{\mathrm{bin}}}\,t_{\mathrm{rel}} \approx 0.33\,t_{\mathrm{rel}},
\end{equation}
assuming a typical binary mass $m_{\mathrm{bin}} \approx 2m_\ast$.
Figure~\ref{fig:trelax} shows the relaxation time, $t_{\rm rel}$, and the dynamical friction timescale, $t_{\rm df}$ (discussed below), as functions of radius for our model. The relaxation time greatly exceeds a Hubble time in the outer regions of the system.

For the bulk of the system ($r \lesssim r_{\rm h}$), this confirms the validity of the collisionless approximation. However, in the innermost regions ($r \lesssim r_{\rm c}$), the local relaxation time is shorter than $13{\,}700,\mathrm{Myr}$, implying that the core experiences approximately two relaxation times by the end of the simulation. As a result, we expect small but non-negligible collisional effects to develop over the course of the simulation, such as mild core contraction or mass segregation in the central regions of the UFD. In particular, binary systems are expected to segregate on the dynamical friction timescale.

\subsection{Binary initial conditions}\label{sec:binic}

Binary pairing is assigned randomly following \citet{weidner2009} and \citet{Wangetal2015}. The logarithm of the semimajor axis is drawn from a uniform logarithmic distribution over the range 0.1--50~AU, ensuring that the primordial population is dominated by hard binaries, following the conclusion on the contribution of hard binaries to velocity dispersion of \cite{2022ApJ...939....3P}. Eccentricities are drawn from a thermal distribution, $f(e)=2e$. Since the 0\% and 50\% models are primarily intended as proof-of-concept cases, we adopted a 30\% binary fraction for most of our plots. This choice is motivated by several considerations: (i) the system is effectively collisionless, with negligible direct stellar collisions (a point that will be demonstrated by our results below); (ii) binary stars play a critical role in shaping the measured velocity dispersion; and (iii) binaries serve as an important proxy for stellar evolution, through processes such as Roche lobe overflow and other binary evolutionary channels. The models are labeled following the nomenclature reported in Table~\ref{tab:ics}.

\subsection{\texttt{NBODY6++GPU}}

High-precision simulations of large stellar systems are performed using \texttt{NBODY6++GPU}, a direct N-body integrator optimized for GPU acceleration \citep{nitadori2012,Wangetal2015} and parallel computing \citep{spurzem2023}. The code is based on the original N-body family developed by Sverre Aarseth, able to integrate systems with up to more than one million particles on modern supercomputing architectures \citep[see, ][]{Wang2016,ArcaSeddaetal2023b, ArcaSeddaetal2024a, ArcaSeddaetal2024b}.

The \texttt{NBODY6++GPU} code uses Kustaanheimo-Stiefel regularization \citep{kustaanheimo1965}, to efficiently handle close binaries. Single and binary stellar evolution is also treated with the "Binary Stellar Evolution" and "Single Stellar Evolution" packages of \cite{hurley2000,hurley2001,hurley2002}, which have been refined and adapted to the current discoveries in stellar evolution \citep[for more information, see][]{Kamlahetal2022a,ArcaSeddaetal2024a,ArcaSeddaetal2023b,ArcaSeddaetal2024b}.
Although \texttt{NBODY6++GPU} is most commonly applied to star clusters, it is well suited for modeling low-density systems such as UFDs. In the present work, DM is not included, but its incorporation through analytic density profiles will be explored in future studies.

\section{Results}\label{results}

\subsection{Dynamical evolution of the UFD}

\begin{figure}
   \centering
   \includegraphics[width=\columnwidth]{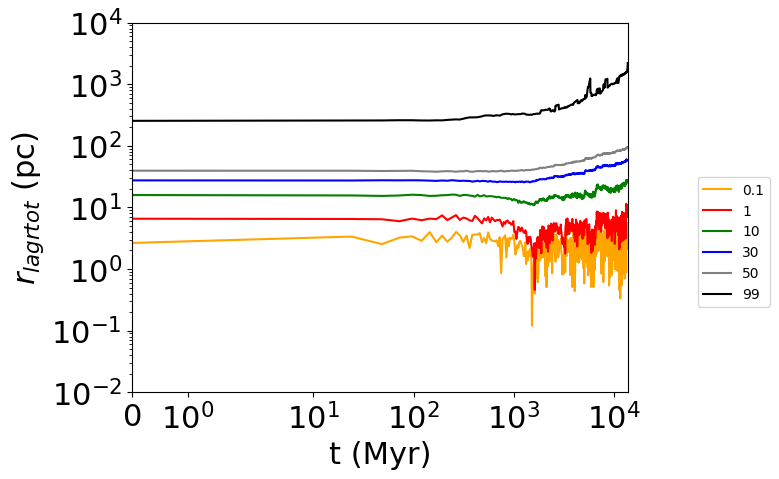}\\
   \includegraphics[width=\columnwidth]{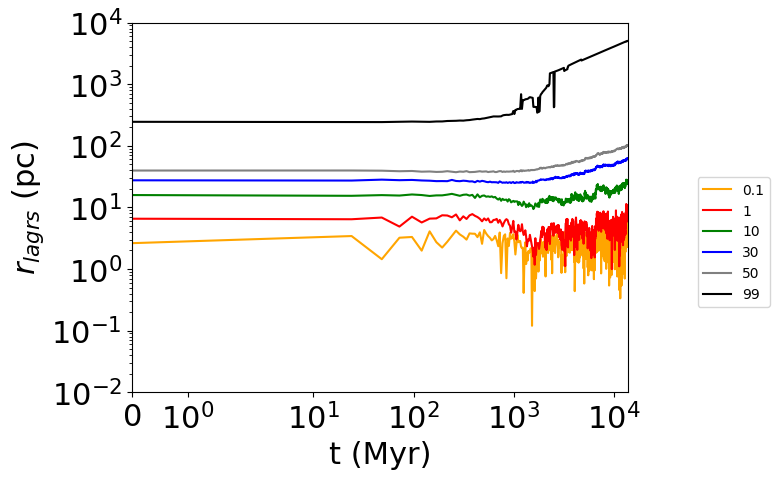}\\
   \includegraphics[width=\columnwidth]{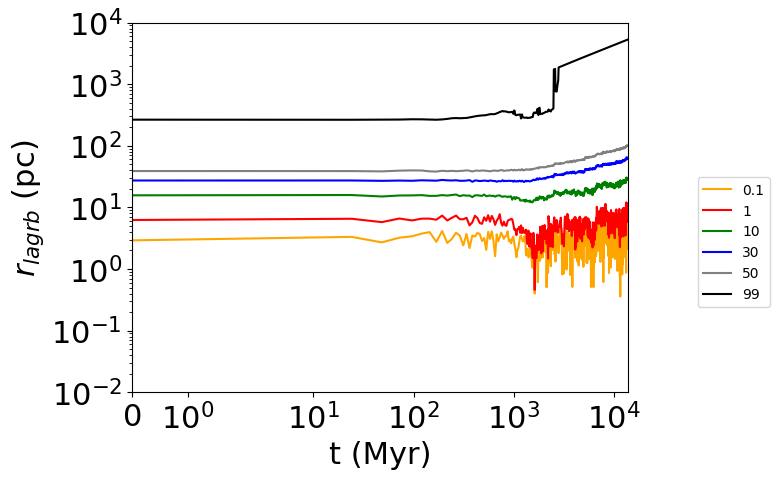}
   \caption{Lagrangian radii evolution of the UFD model U30 over one Hubble time, using the time-updated value of the total mass. Top: For both single and binary stars. Middle: For single stars. Bottom: For binary stars.
              \label{fig:lagr}}
    \end{figure}

The evolution of the Lagrangian radii of the UFD shown in Fig.~\ref{fig:lagr} illustrates the behavior of both the single-star and binary components. With the exception of model U0 -- which contains no binaries -- all models exhibit a similar evolutionary trend in their binary populations. In particular, in the innermost regions, models with lower binary fractions show slightly earlier mass segregation, since binaries tend to delay core collapse (see Appendix ~\ref{sec:LR02040}, where we display and compare the Lagrangian radii of models U0, U20, and U40).

Figure~\ref{fig:lagr} highlights the nearly collisionless nature of UFDs. Although the system loses mass from its most massive stars -- which evolve into black holes or neutron stars within the first 100 Myr -- the Lagrangian radii are only mildly affected, and only in the innermost regions. This behavior is a consequence of the low stellar density, even at small radii, which results in rare stellar encounters. The outer regions also expand slowly, since the UFD is treated as an isolated system.

Although the cluster is considered isolated, we adopted a mass-loss criterion defined by
$r_{\rm esc} = 20\, r_{\rm scale}$,
where $r_{\rm scale}$ is the scale radius of the system, which we set equal to the half-mass radius. Thus, if the UFD expands to 100 pc, the escape radius becomes 2{\,}000 pc. All stars that move beyond $r_{\rm esc}$ are considered lost from the system.

The UFD does not show any significant structural variation until approximately 3{\,}000 Myr, when bumps appear in the inner Lagrangian radii and the outer shells expand. This behavior is a consequence of the onset of mass segregation, which increases the encounter rate in the central regions and triggers a mild core contraction. As stellar evolution proceeds, white dwarfs form, and the associated mass loss halts further core contraction once the majority of red giants have evolved into white dwarfs.

We therefore conclude that, prior to the onset of mass segregation, binary disruptions and stellar ejections are limited to processes related to binary evolution (e.g., Roche lobe overflow). This outcome is also a consequence of our choice of initially hard binaries, which is well motivated by the results of \cite{2022ApJ...939....3P} and appropriate for the scope of this study.

The middle panel of Fig.~\ref{fig:lagr} shows the evolution of the Lagrangian radii for the single-star population of model U30. Since single stars constitute 70\% of the total stellar population in this model, they play a dominant role in determining the long-term evolution of the UFD, which remains largely collisionless. After 3{\,}000 Myr, the positions of single stars are more strongly perturbed in the inner regions as a result of mass segregation, which drives lower-mass objects toward larger radii.

The bottom panel of Fig.~\ref{fig:lagr} shows a similar pattern for the Lagrangian radii of the binary population. Both single stars and binaries are driven toward the outer regions, but the effect is more pronounced for binaries. In particular, the 99\% Lagrangian radius of the binary component increases rapidly, whereas the corresponding single-star shell expands more steadily. This behavior is induced by mass segregation and core contraction, which preferentially drive low-mass single stars and low-mass binaries outward. As a result, a prominent expansion peak appears around the time when the system ceases segregating, at approximately 5 Gyr for model U30.

By 3 Gyr, the UFD has already expanded beyond 1{\,}000 pc, a scale at which tidal stripping by the MW would become significant in a more realistic, non-isolated scenario. However, since the UFD is modeled as isolated in our simulations, it does not experience evaporation in the denser regions until several relaxation times have elapsed, well beyond a Hubble time.

\begin{figure}
   \centering
   \includegraphics[width=\columnwidth]{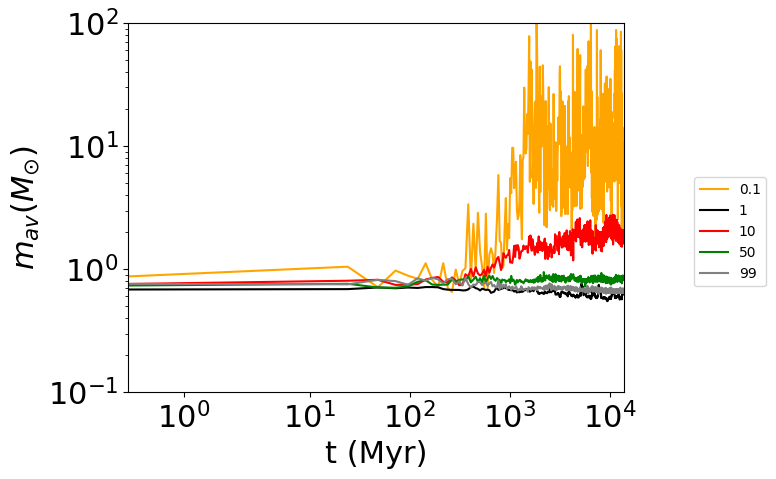}\\
   \caption{Average mass within the different Lagrangian radii of the U30 model over one Hubble time.
              \label{fig:masssegr}}
    \end{figure}

As discussed above, the key feature revealed by these plots is the presence of mass segregation, which is quantified by the average stellar mass within the Lagrangian shells and shown in the top panel of Fig.~\ref{fig:masssegr}. Mass segregation remains negligible up to approximately 1 Gyr, after which a pronounced segregation of higher-mass stars develops in the central regions. Nevertheless, its impact on the overall dynamical evolution of the inner regions is limited, owing to the large spatial extent of the UFD at the time segregation becomes effective, which results in long segregation timescales.

The formation of new binaries -- amounting to only $\sim 0.004\%$ of the total stellar population -- does not significantly affect the evolution of the Lagrangian radii of either the single-star or binary components. As we show in the next section, these newly formed systems are wide binaries that are rapidly disrupted (within $\lesssim 3$ Myr), rendering their contribution to the global dynamical and luminosity evolution of the UFD entirely negligible.

\subsection{Mass loss and binaries lost in the UFD}

\begin{figure}
   \centering
  \includegraphics[width=0.9\columnwidth]{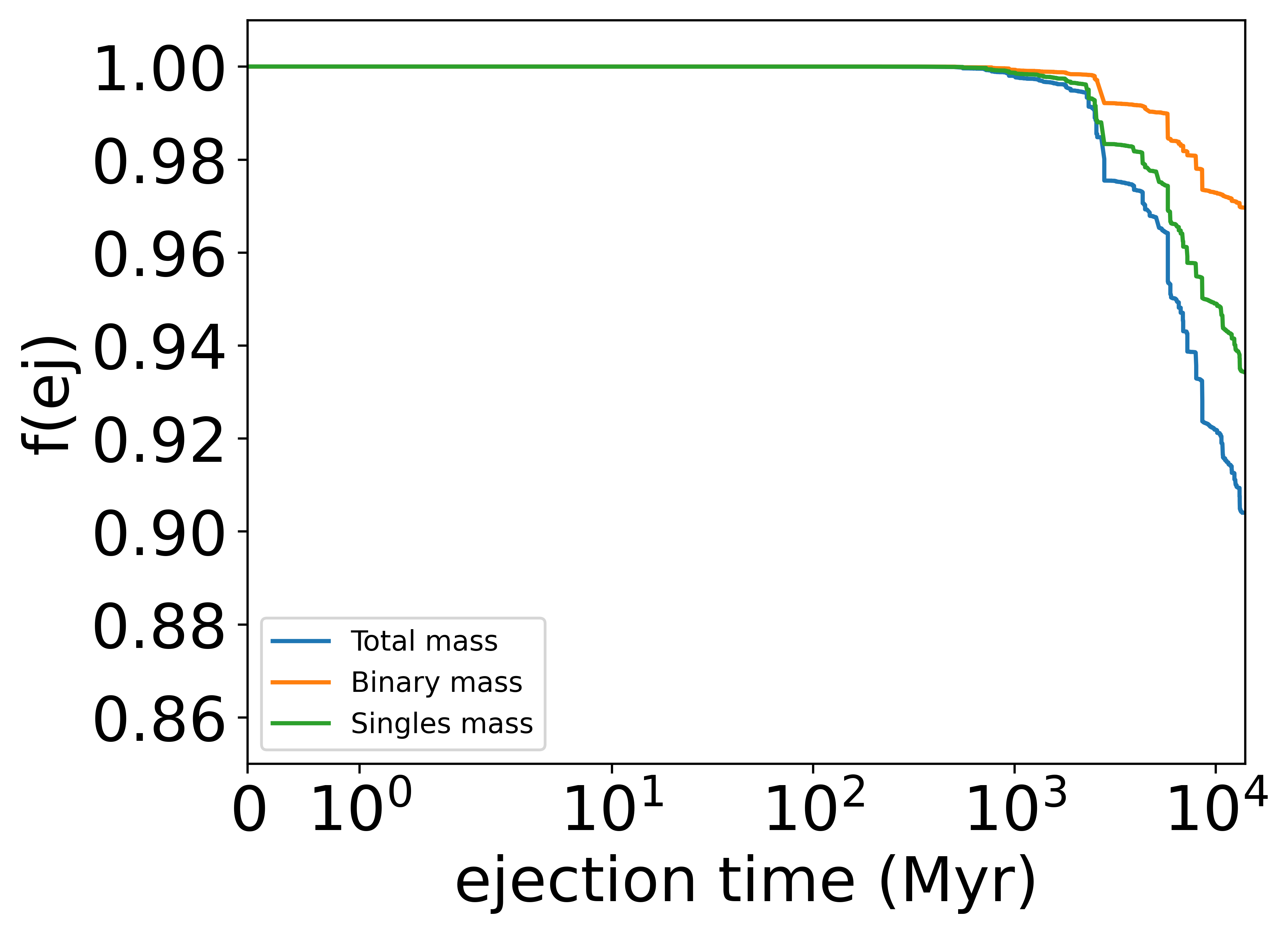}\\
   \includegraphics[width=\columnwidth]{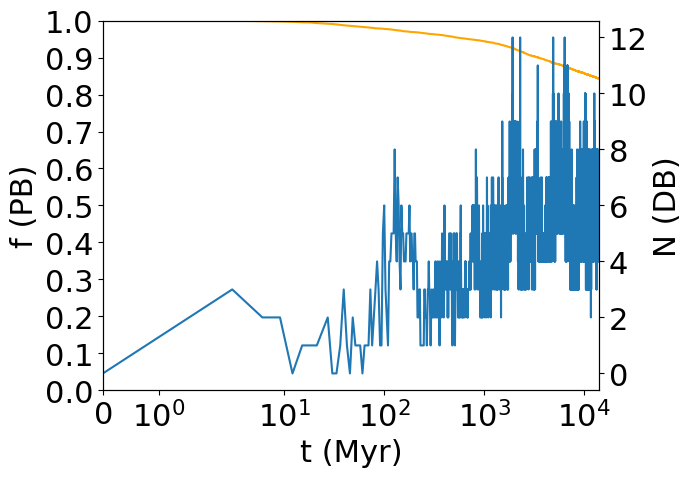}
   \caption{Top: Fraction of mass present in the UFD, compared to the total mass of each component over the simulation time. 
   Bottom: Evolution of the members of  dynamical binaries the survive in the UFD (blue, right-side ordinate) and the fraction of primordial binaries that survive in the UFD (orange, left-side ordinate). 
             \label{fig:ejected}}
   \end{figure}

The total mass ejected is extremely small compared to that of denser systems, as shown in the top panel of Fig.~\ref{fig:ejected}. The total mass lost amounts to 9\% of the initial mass. Notably, this loss is driven primarily by stellar evolution ($\sim 6\%$, corresponding to 66\% of the total mass loss) and, to a lesser extent, by ejections ($\sim 3\%$, corresponding to 33\% of the total mass loss). Approximately 6\% of the mass loss originates from single stars, while the remaining 3\% comes from binary systems, with percentages computed relative to the initial total mass of the UFD.

The first episode of mass loss in the UFD occurs during the early evolutionary phase, when the most massive stars rapidly evolve into neutron stars or black holes within the first $\sim 25$--100 Myr. In star cluster systems, neutron stars are typically ejected \citep[e.g.,][]{flammini2025}; however, the large spatial extent of the UFD results in a negligible amount of mass loss associated with neutron-star ejections, which instead occur much later in the evolution (see the next section). This behavior is also a consequence of treating the system as isolated in our models.

As noted above, less massive objects tend to migrate toward the outer regions due to two-body relaxation after approximately 3{\,}000 Myr. In the bottom panel of Fig.~\ref{fig:ejected}, we show the fraction of disrupted and ejected primordial binaries, together with the absolute number of dynamically formed binaries. The majority of dynamically formed binaries are extremely wide systems (with semimajor axes $a > 10^3$ AU) and are therefore disrupted on short timescales compared to hard primordial binaries. As is well known, most dynamically formed binaries become bound at large separations from their host stars. In our simulations, these dynamically formed binaries are consequently very short-lived.

As a result, the dynamical binary population contributes negligibly to both the luminosity and the velocity dispersion of the system, accounting for only $\sim 0.004\%$ of the total number of binaries. In contrast, primordial binaries are lost at the $\sim 10\%$ level before 3{\,}000 Myr. Most of these losses occur within the first 100 Myr due to stellar evolution, particularly the formation of compact objects, which can unbind the binary system. The remaining $\sim 6\%$ of primordial binaries are lost between 3{\,}000 Myr and a Hubble time. As the UFD continues to expand, stellar crossing times increase significantly, making further ejections progressively less likely.

By a Hubble time, the UFD has lost approximately 16\% of its original primordial binary population. Of these lost binaries, $\sim 3\%$ are removed through ejections, while $\sim 13\%$ are lost through binary disruption.

Finally, we examined the evolution of the semimajor axis of binaries in model U30, both including dynamically formed binaries (shown in blue) and considering only primordial binaries. The average semimajor axis in both cases remains similar during the first $\sim 100$ Myr, after which it gradually increases. This evolution is driven by stellar evolution, particularly the formation of white dwarfs and giant stars, which can significantly alter binary orbital parameters. The peak observed at the end of the curve is caused by the inclusion of dynamically formed binaries; when these systems are excluded, the evolution of the semimajor axis is considerably smoother.


  \begin{figure}
   \centering
  \includegraphics[width=\columnwidth]{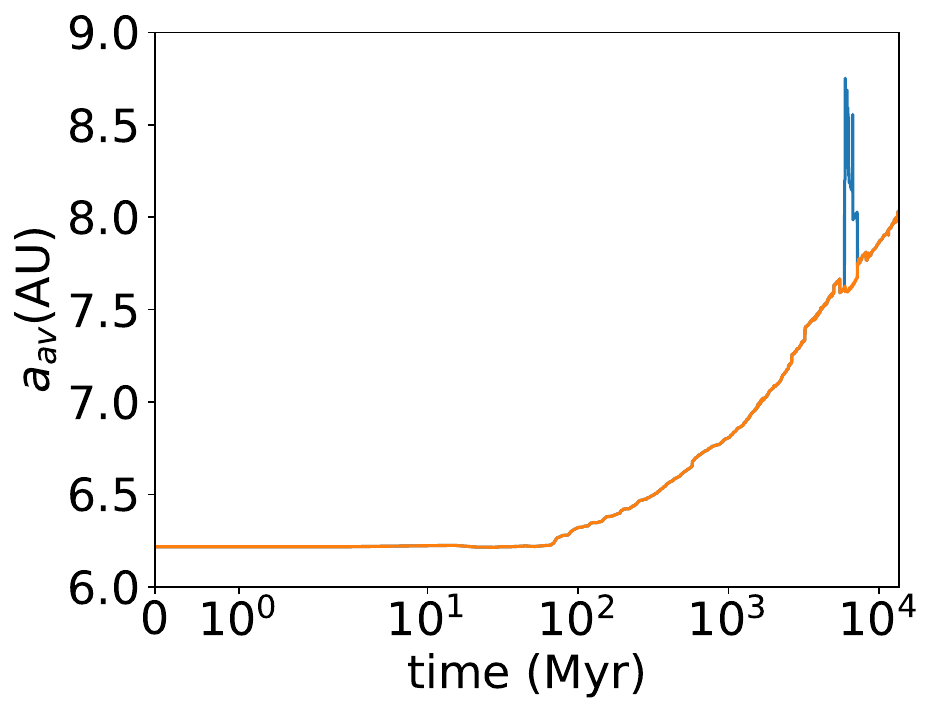}\\
   \caption{Evolution of the average semimajor axis for the binaries in model U30. The orange curve corresponds to the sample of primordial binaries; the blue curve also takes the dynamical binaries into account.  \label{fig:smaaverage}}
   \end{figure}

\subsection{Impact of stellar evolution}

\begin{table}
        \centering
        \caption{Different evolutionary stages of stars at different times.
    }
    \label{table:SEV}   
    \begin{tabular}{llllll} 
                \hline
Time (Myr) &   MS   &     GS   &   WD   &  NS   &    BH \\
\hline\hline
0    &   100.00 &  0.00   &       0.00    &   0.00    &    0.00  \\
25   &   99.50 &  0.04  &        0.00    &   0.19  &    0.27\\
100  &   99.40 &  0.08  &        0.13  &   0.14  &    0.25\\
1\,000  &  96.40 &  0.44 &        2.99 &  0.02   &    0.15\\
5\,000  &  92.12 &  0.26 &         7.46 &   0.01    &    0.15\\
13\,700   & 88.22 &  0.18 &      11.45 &   0.01    &    0.14 \\
                \hline\hline
        \end{tabular}
\tablefoot{    Percentage of stars in different evolutionary stages at different key times in the simulation U30. \\MS = main sequence stars; GS = red giant and asymptotic giant branch stars; WD = white dwarfs; NS = neutron stars; BH = black holes. }
\end{table}

   \begin{figure}
   \centering
   \includegraphics[width=1\columnwidth]{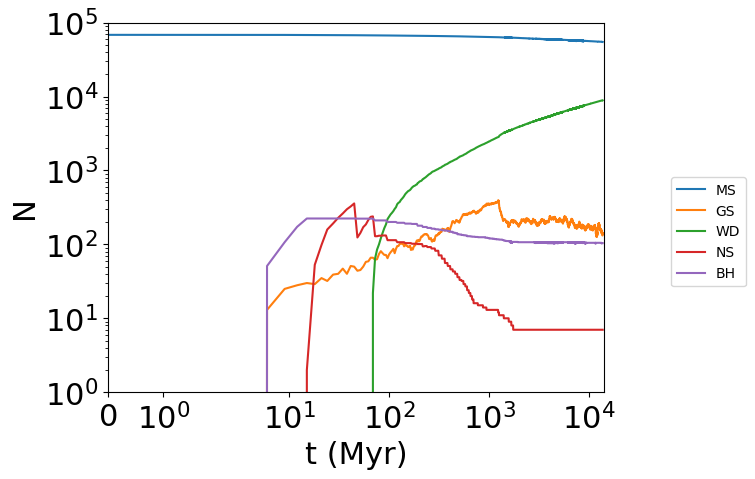}\\
      \caption{Number of stars in different evolutionary stages over time (see Table \ref{table:SEV}).
    }
         \label{fig:Ntypeovertime}
   \end{figure}
   \begin{figure}
   \centering
   \includegraphics[width=1\columnwidth]{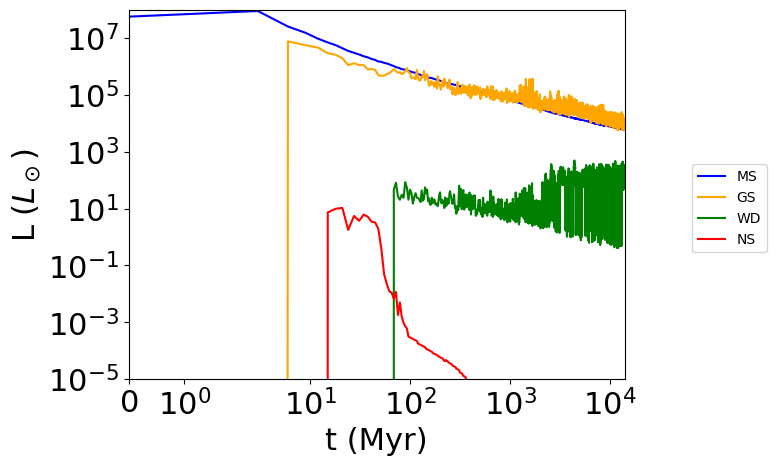}\\
      \caption{Total bolometric luminosity of stars in different evolutionary stages over time (model U30; see Table \ref{table:SEV}).}
         \label{fig:Ltypeovertime}
   \end{figure}
      \begin{figure}
   \centering
   \includegraphics[width=\columnwidth]{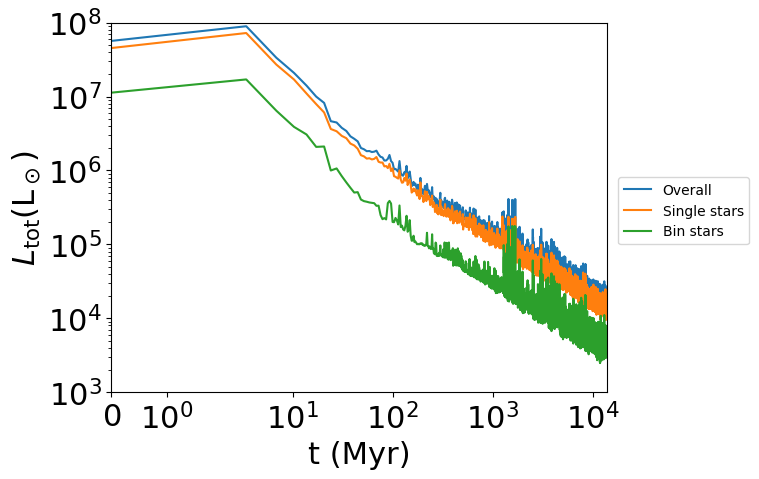}\\
   \caption{Time evolution of the total  bolometric luminosity (blue) of model U30 and of the luminosity of single stars (orange) and binaries (green). 
        \label{fig:thermodynamics}}
    \end{figure}

The role of stellar evolution is extremely important in a nearly collisionless system. For this reason, a detailed examination of the stellar population content of the simulated system is particularly informative. Table~\ref{table:SEV} reports the fraction (in percent) of the different stellar types populating the U30 UFD model over its evolution, from $t=0$ to one Hubble time. The fraction of helium stars is sufficiently small that their overall contribution is negligible; consequently, they are not included in Table~\ref{table:SEV}.

We considered the following specific evolutionary timestamps:
\begin{enumerate}
    \item The 25 Myr timestamp is relevant because many massive stars begin rapid evolution and become neutron stars or black holes;
    \item The 100 Myr timestamp corresponds to the formation of the first white dwarfs, primarily in binary systems;
    \item The 1{\,}000 Myr timestamp approximately marks the onset of mass segregation, when the average stellar mass in the inner regions begins to increase (see Fig.~\ref{fig:masssegr});
    \item The 5{\,}000 Myr timestamp corresponds to a phase in which the number of white dwarfs has nearly doubled relative to the previous timestamp;
    \item The 13{\,}700 Myr timestamp marks the end of the simulation, corresponding to one Hubble time.
\end{enumerate}

At a Hubble time, white dwarfs constitute approximately 11\% of the initial total number of stars and about 13\% of the stellar population remaining at the end of the simulation. In Fig.~\ref{fig:Ntypeovertime} we show the number of stars of each stellar type as a function of time, providing an overview of the stellar population evolution in the UFD. One of the most prominent features is the peak in the formation of giant stars and the rapid increase in the number of white dwarfs, which become the dominant compact object population after $\sim 100$ Myr. Both of these characteristics were first identified in dwarf galaxies by \cite{rcd1980}.

Black holes are largely retained by the galaxy, whereas neutron stars escape in significant numbers due to supernova natal kicks \citep[see stellar evolution recipe C in][]{Kamlahetal2022a}. However, neutron-star escape occurs predominantly at later times, which can be explained by the crossing timescale of the system, $t_{\rm cr} \approx 62$ Myr. The fastest neutron stars escape first, followed by progressively slower ones. White dwarfs, by contrast, are almost entirely retained by the UFD.

In Fig.~\ref{HRtimes} we show the Hertzsprung--Russell diagram of the U30 model in the $L_{\rm bol}$--$T_{\rm eff}$ plane at several key evolutionary times, including 2 Gyr to capture the system during the onset of mass segregation. The HR diagrams are generated using the stellar and binary evolution prescriptions of \cite{Hurley2005} and their updated implementations \citep{Kamlahetal2022b, spurzem2023, vergara2025}.

Figure~\ref{fig:Ltypeovertime} shows the luminosity evolution of the different stellar populations in the UFD. Black holes are excluded, as their luminosity contribution is negligible. The most massive stars evolve into neutron stars or black holes within the first 20--25 Myr, forming compact remnants with low luminosities. Main-sequence stars dominate the luminosity budget during the first 100 Myr, while red giant branch stars become comparable contributors at $\sim 1{\,}000$ Myr and subsequently dominate the luminosity of the UFD. This occurs as stars near the upper end of the main sequence ignite hydrogen shell burning and evolve into red giants, leading to a substantial increase in the total luminosity of the system.

Blue stragglers are extremely rare in our simulations, with only a few forming, consistent with previous observational findings \citep{momany2015}. The luminosity contribution of neutron stars remains entirely negligible, even at their peak luminosities of $\sim 10^1\,L_{\odot}$. By the end of the simulation, white dwarfs constitute the third most significant contributors to the total luminosity. Since not all main-sequence stars have evolved into white dwarfs by a Hubble time, main-sequence stars continue to contribute more strongly than white dwarfs even at late times.

In summary, the luminosity of the UFD is ultimately dominated by red giant stars. As the system begins to segregate, an increasing number of red giants form, enhancing their contribution to the total luminosity. At this stage, red giant stars become approximately five times more luminous than main-sequence stars, significantly shaping the integrated light of the UFD.

\subsection{Velocity dispersions of UFD models}
   \begin{figure}
   \centering
   \includegraphics[width=\columnwidth]{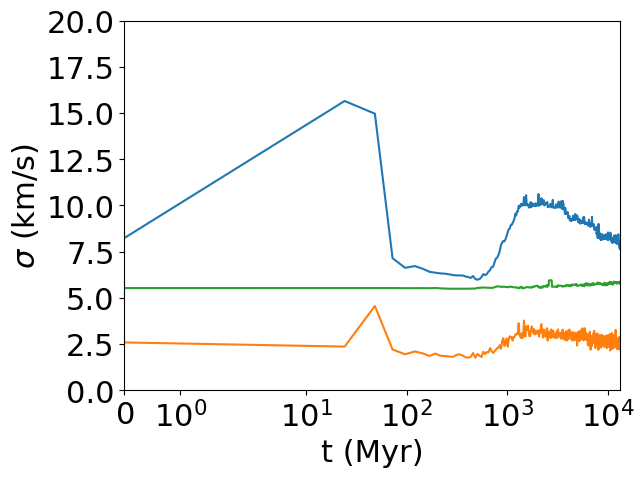}\\
   \includegraphics[width=\columnwidth]{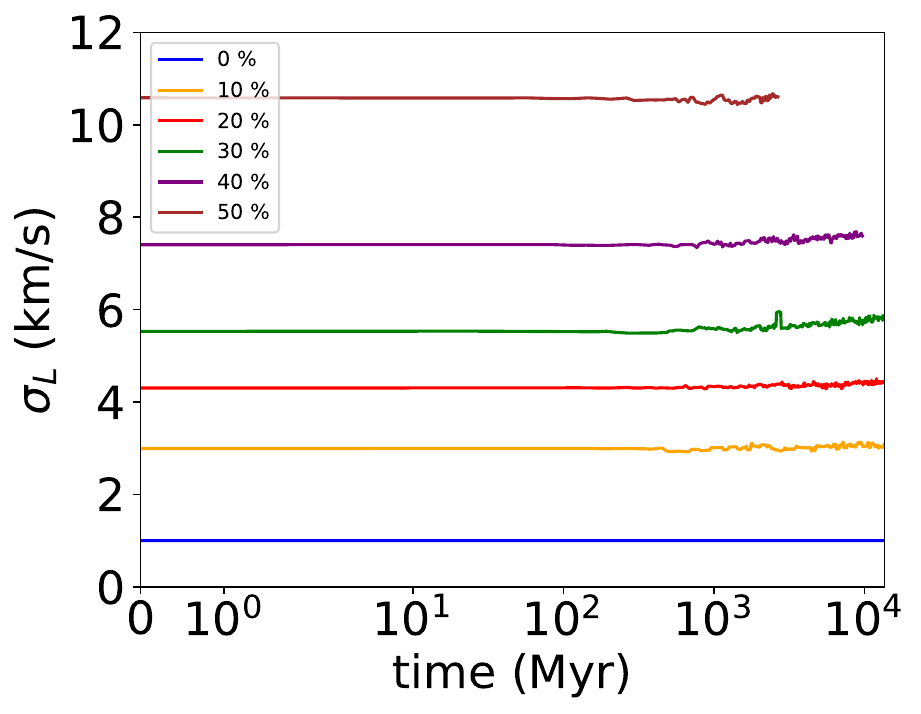}
   \caption{Model U30. Top:  $\sigma_{\rm g}$ (blue), $\sigma_{\rm g,cm}$ (orange), and  $\sigma_{\rm L}$ (green) evolution over time.
   Bottom: $\sigma_{\rm L}$ for models with initial binary fractions in the $0-50\%$ range.
              \label{fig:dispersioni}}
    \end{figure}

\begin{figure*}
\centering
\resizebox{\textwidth}{!}{\begin{tabular}{|c|c|c|}
\hline
\subfloat{\includegraphics[width=60mm]{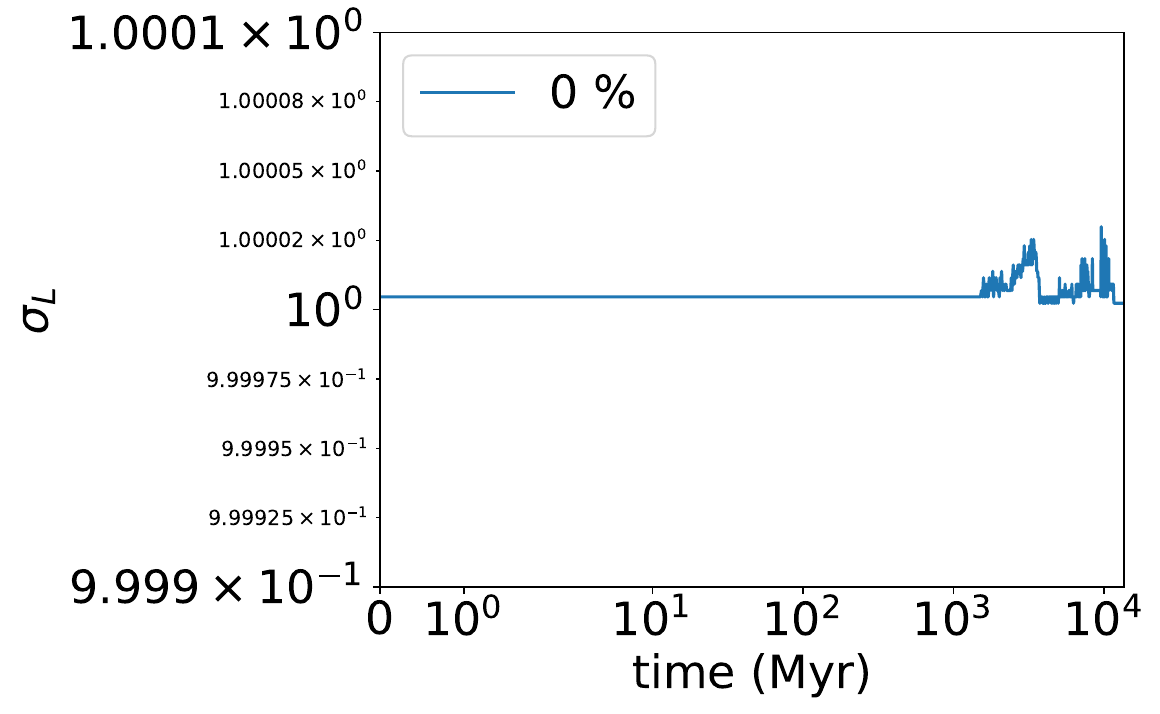}}
     {}
&
\subfloat{\includegraphics[width=60mm]{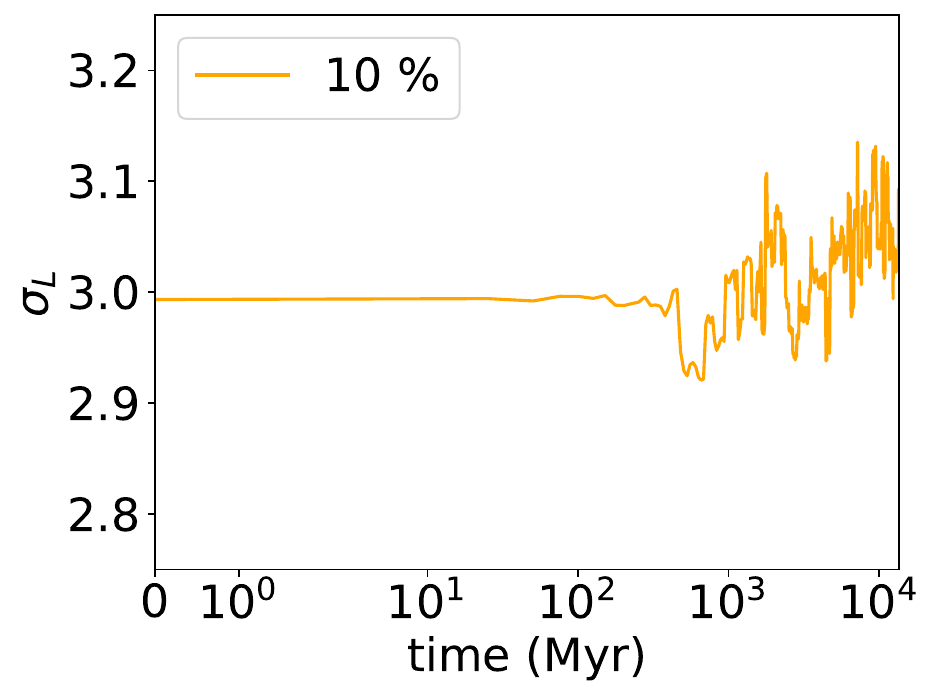}}
     {}
&
\subfloat{\includegraphics[width=60mm]{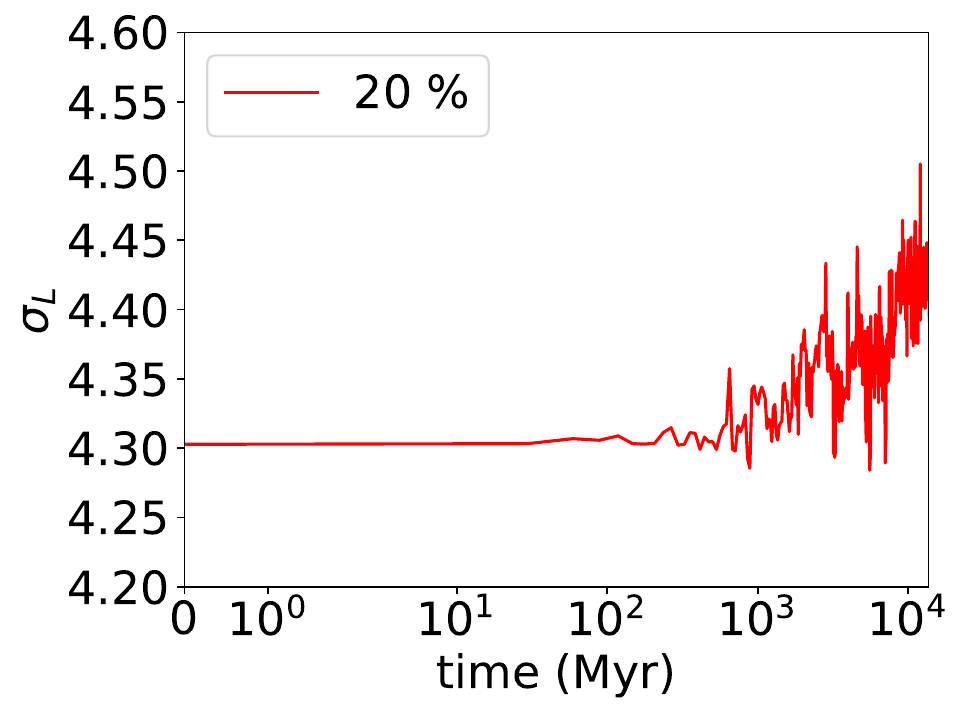}}
     {}
     \\
     \hline
\subfloat{\includegraphics[width=60mm]{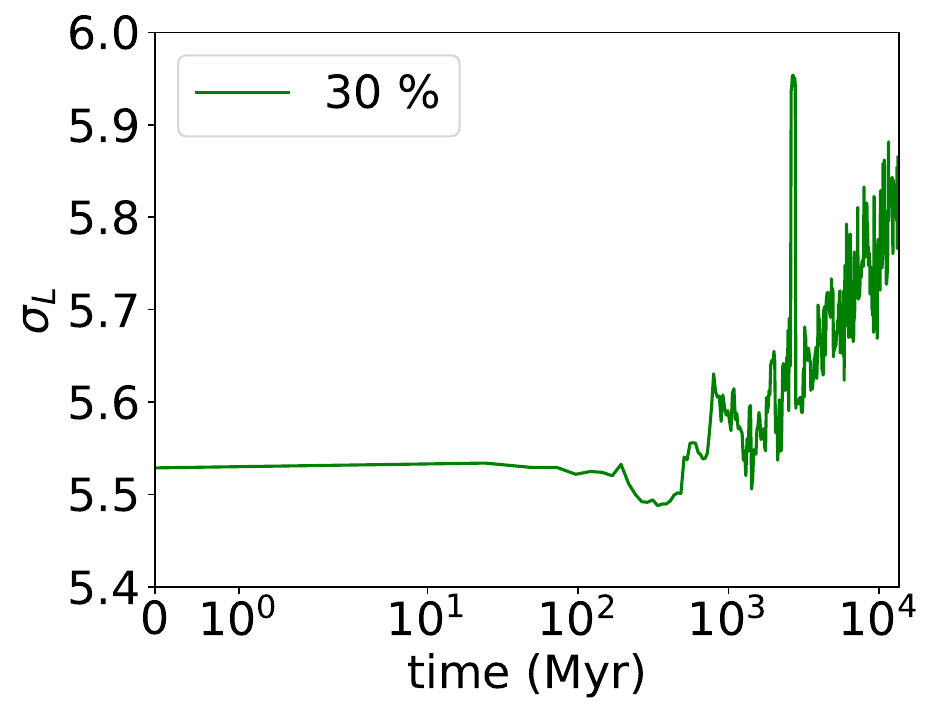}}
     {}
     &
     \subfloat{\includegraphics[width=60mm]{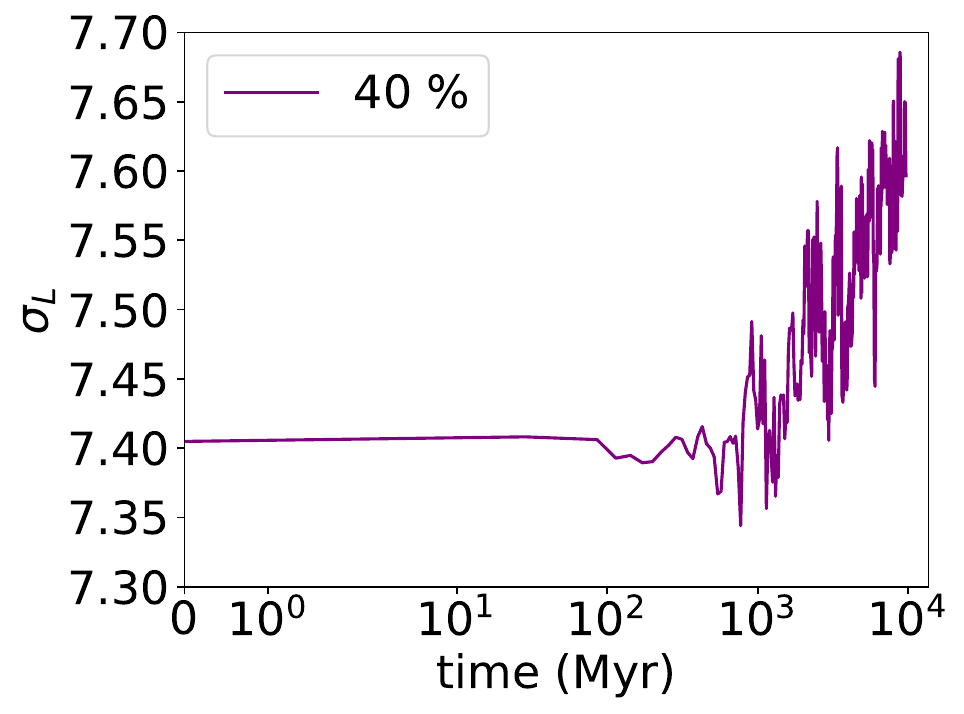}}
     {}
&
\subfloat{\includegraphics[width=60mm]{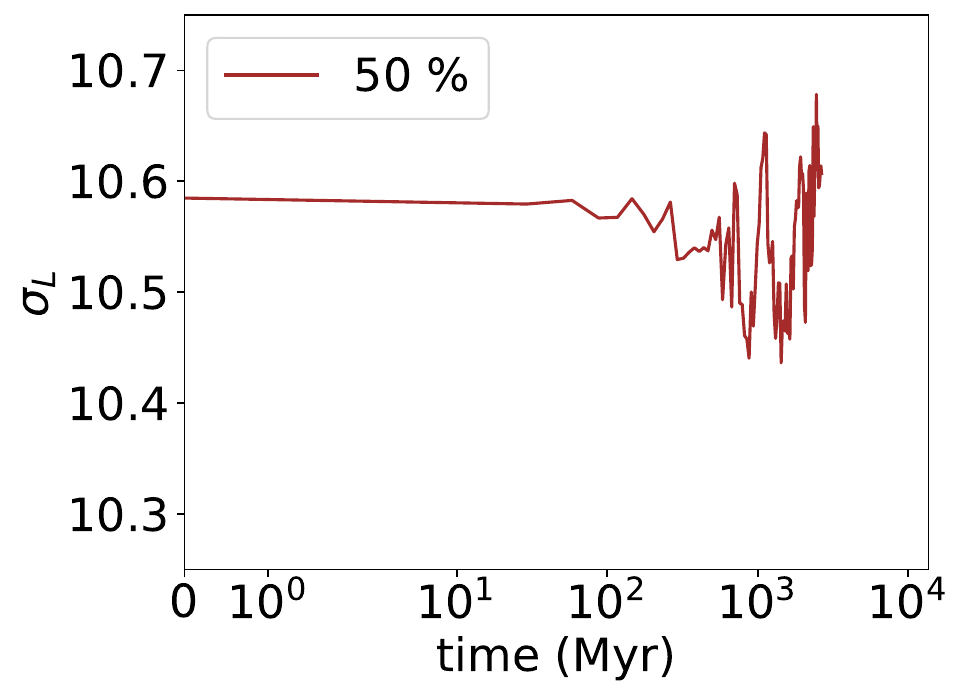}}
     {}
\\
\hline
\end{tabular}}
\caption{Velocity dispersions ($\sigma_{\rm L}$) for all models, from U0 (top left)   to U50 (bottom right).}
\label{fig:zoomsigmal}
\end{figure*}

As anticipated in the Introduction, one of the primary motivations of this study is to assess the role of binary stars in estimating the virial mass of low-mass stellar systems, such as dSphs and UFDs. In this way, observational measurements of this fundamental quantity can be more accurately interpreted. Although the effect of binaries in inflating the velocity dispersion of stellar systems has already been explored analytically (see, e.g., \citealt{2022ApJ...939....3P}), in this work we present, for the first time, a fully self-consistent dynamical N-body model of a UFD-sized galaxy. 

In particular, we simulated the evolution of a UFD-like system from the initial burst of star formation up to a Hubble time, adopting a $1{:}1$ star--particle representation. Compared to previous studies, this approach allows us to follow the complete dynamical evolution of the system while treating binary stars as living components of the stellar population, rather than as a static or statistical correction.

To evaluate the velocity dispersion -- which represents both the true kinetic support against gravity and the spurious contribution from binary orbital motions -- we considered different prescriptions for computing velocity dispersions in our simulated UFDs. We refer the reader to \cite{2020ApJ...896..152R} and \cite{2022ApJ...939....3P} for a detailed discussion of these approaches, which are designed to reproduce, with varying degrees of realism, quantities comparable to those measured in spectroscopic observations. In this study, we adopted both a direct evaluation of the velocity dispersion and a luminosity-weighted one.

To disentangle the contribution from binary orbital motion, we compared: (i) the true virial velocity dispersion, which provides the kinetic support against gravitational contraction and, for binaries, accounts only for the center-of-mass motion (thereby excluding internal orbital motion); and (ii) a blind determination, in which all stars are treated as single objects, implicitly including the contribution from internal binary motions. The latter clearly represents an overestimate of the true kinetic support against gravity. For the former, we also included the binary contribution from the weighted luminosity distribution of binaries.
We can describe the classical equation for velocity dispersion as

\begin{equation}
\sigma_{\text{g}} = \sqrt{\frac{\sum_{i=1}^{N} (v_i - \langle v \rangle)^2}{N}}   \quad , 
\end{equation}

\noindent where $i$ is the single star component, while  $\langle v \rangle$ is the average velocity of all the stars. We treated each star as a single star, marking the blind approach. In order to analyze the binary center of mass and luminosity weighted approach, we changed the factor $v_i$ for the stars in binaries.
For the center of mass binaries, we wrote this as
\begin{equation}
v_{\mathrm{cm},\mathrm{j}} = \frac{m_{A,j} \, v_{A,j} + m_{B,j} \, v_{B,j}}{M_j}\quad ,
\end{equation}
where A and B are the binary components indexes, while $M_j = m_A,j + m_B,j$ is the sum of the two masses. A similar approach is done for the luminosities:\begin{equation}
    v_{\mathrm{L},\mathrm{j}} = \frac{L_{A,j} \, v_{A,j} + L_{B,j} \, v_{B,j}}{L_j}\quad ,
\end{equation}
where instead of the mass, we use the luminosity. Thus, we obtain
\begin{equation}
\sigma_{\mathrm{g,cm}} = \sqrt{\frac{1}{N_s + N_b} \left( \sum_{i=1}^{N_s} v_i^2 + \sum_{j=1}^{N_b} v_{\mathrm{cm},j}^2 \right)}\quad ,
\end{equation}
where $N_\textrm{b}$ is the number of binaries. Similarly for the luminosity-weighted case:
\begin{equation}
\sigma_{\mathrm{L}} = \sqrt{\frac{1}{N_s + N_b} \left( \sum_{i=1}^{N_s} v_i^2 + \sum_{j=1}^{N_b} v_{\mathrm{L},j}^2 \right)}\quad .
\end{equation}
The early evolution of both $\sigma_{\rm g}$ and $\sigma_{\rm g,cm}$ in the top panel of  Fig.~\ref{fig:dispersioni} is dominated by stellar evolution. The most massive stars evolve into neutron stars and black holes within the first 20--25 Myr. The natal kicks imparted to neutron stars significantly increase their velocities, producing a sharp rise in the velocity dispersion. This effect is visible in both $\sigma_{\rm g}$ and $\sigma_{\rm g,cm}$, since both measures are sensitive to changes in stellar velocities.

At later times, around 3{,}000 Myr, the velocity dispersion increases again as a consequence of core contraction associated with the onset of mass segregation. The increased encounter rate in the central regions raises the velocities of stellar members, leading to a modest enhancement of the dispersion.

In contrast, the luminosity-weighted velocity dispersion, $\sigma_{\rm L}$, is much less sensitive to mass loss and remains nearly constant throughout the simulation, aside from a mild perturbation around 3{,}000 Myr. This behavior reflects the dominant contribution of red giant branch stars to the total luminosity during this phase. Once the number of red giants declines, $\sigma_{\rm L}$ returns to its previous value. Overall, $\sigma_{\rm L}$ is effectively time-independent, with variations limited to at most $\sim 1\%$ of its initial value. In model U0, which contains no primordial binaries and only a negligible number of dynamically formed binaries, variations in $\sigma_{\rm L}$ are correspondingly minimal.

In the bottom panel of Fig.~\ref{fig:dispersioni}, the velocity dispersion remains approximately constant in time for all models, consistent with the largely collisionless nature of the system and the relatively stable binary fraction over the simulation.

The simulation of model U50 was not completed to a full Hubble time due to computational constraints on the \textsc{Leonardo} booster system. However, this does not affect our conclusions, as the velocity dispersion in this model clearly reaches a stable plateau, consistent with the behavior observed in the fully evolved models.

As expected, the velocity dispersion increases significantly with increasing binary fraction, reaching values up to $\sim 10.58$ times larger than those measured in the absence of binaries. The data shown in Fig.~\ref{fig:dispersioni} confirm the expected approximately linear dependence of $\sigma^2$ on the binary fraction $f_b$. The best-fitting linear relation, which remains nearly unchanged over time, is (for $f_b \geq 0.1$)

\begin{equation}
\sigma_{\rm L}^2 = 201.39 f_b - 12.69 \; \mathrm{km}^2\,\mathrm{s}^{-2},
\end{equation}

\noindent with an rms scatter of 14.80. A better fit, still for $f_B \ge 0.1$, is obtained with a quadratic formula

\begin{equation}
\sigma_{\rm L}^2 = 768.0f_b^2 - 218.321 f_b+26.0 \; \mathrm{km}^2\,\mathrm{s}^{-2},
\end{equation}
giving an rms scatter of 4.345. Since the virial mass estimate scales with the square of the velocity dispersion, this implies an overestimate of the virial mass by a factor of $\sim 27$.
For completeness, we also present a zoomed-in view of $\sigma_{\rm L}$ for each model in Fig.~\ref{fig:zoomsigmal}. The variations are small in all cases, consistent with the collisionless character of the UFD. In model U0, the absence of binaries leads to only a minor change in $\sigma_{\rm L}$ during the onset of mass segregation. This behavior further reflects the modest overall mass loss experienced by the system, confirming that models with lower numerical noise provide an accurate representation of UFD dynamics.
Finally, a visual comparison with Fig. 2 of \cite{sim19} of ours models is present in Appendix C. Our models covers the observed range from \cite{sim19} in the velocity dispersion quite well. Note that our simulated UFDs are on the high luminosity edge of what \cite{sim19} classified as UFDs (dwarfs with $M_V >-7$).

\section{Conclusions}\label{conclusions}

We computed six full N-body models representing UFDs populated by both single and binary stars, spanning a range of binary fractions, as described in Sect.~\ref{sec:binic}. The main characteristics of the models are summarized in Table~\ref{tab:ics}. Each model was evolved for a Hubble time following the initial burst of star formation, which allowed us to follow the complete dynamical evolution of the systems.

Ultra-faint dwarf galaxies  are generally expected to be quasi-collisionless systems. We consistently find a broadly similar global dynamical evolution across all models, with differences arising primarily from the varying binary fractions and from stellar evolution. In particular, the luminosity-weighted velocity dispersion -- which is the quantity most relevant for observational mass estimates of dwarf galaxies -- is significantly affected by the fraction of binaries present in the system.

In the following, we summarize the main conclusions of this study:

\begin{enumerate}
    \item The long-term evolution of a UFD can be divided into two main phases. During the first phase, lasting until approximately 3{,}000 Myr, the system remains nearly stationary. In this early stage, some neutron stars and black holes are ejected following supernova explosions, but these events have only a minor impact on the global dynamical evolution. Afterward, the system shows a weak collisional behavior, reaching partial relaxation at around 5{,}000 Myr. Mass segregation develops primarily in the most compact central regions (within $\sim 10$ pc).
    
    \item The role of binaries in the long-term evolution of the system is mainly to moderate the contraction of the inner regions. Larger binary fractions tend to delay the onset of mass segregation, leading to a slightly more extended core structure.
    
    \item The final state of the simulated systems is characterized by a luminosity budget dominated by red giant stars, in agreement with observational evidence for UFDs \citep{sim19}, although they make up, in number,  less than $1 \%$ of the total.
    
    \item With our detailed numerical simulations, we confirm earlier statistical indications from \cite{2022ApJ...939....3P} that the presence of unresolved binaries significantly inflates the estimates of the system velocity dispersion, thereby leading to a substantial overestimate of the virial mass. These results reduce the need to invoke large amounts of DM in very small, low-luminosity systems such as UFDs.
\end{enumerate}

In conclusion, our study demonstrates that UFDs are not perfectly collisionless systems and that their dynamical evolution is nontrivial, being partly governed by their binary content, which remains significant even after a Hubble time. Moreover, we validate previous statistical estimates of the impact of internal binary motions on observational determinations of galaxy velocity dispersions. In particular, sufficiently hard binaries survive throughout the evolution of the galaxy, and their rapid internal motions effectively inflate the Doppler-derived velocity dispersion. The next step of this research would be to include a possible halo of DM, to test its role in the overall dynamics of these small galaxies.

\begin{acknowledgements} 
This work was possible thanks to the use of the CINECA Leonardo booster, in the frame of the project "The Case of Dark Matter Overabundance in Small Galaxies", which was awarded 80,000 core hours.
This material is based upon work supported by Tamkeen under the NYU Abu Dhabi Research Institute grant CASS. 
R.S. acknowledges support from NAOC International Cooperation Office in 2023, 2024, and 2025, Chinese Academy of Sciences President's International Fellowship Initiative for Visiting Scientists (PIFI, 2026PVA0089), and the National Natural Science Foundation of China (NSFC) under grant No. 12473017.
\end{acknowledgements}

\bibliographystyle{aa}
\bibliography{bibliography}{}

\begin{appendix}

\section{Lagrangian radii of models U0, U20, and U40}\label{sec:LR02040}

The figure represents the time evolution of the Lagrangian radii containing from $0.1 \%$ to $99 \%$ of the actual total mass for models U0, U20, and U40.

\begin{figure}[!ht]
\centering
\resizebox{\textwidth}{!}{\begin{tabular}{|c|c|c|}
\hline
\subfloat{\includegraphics[width=60mm]{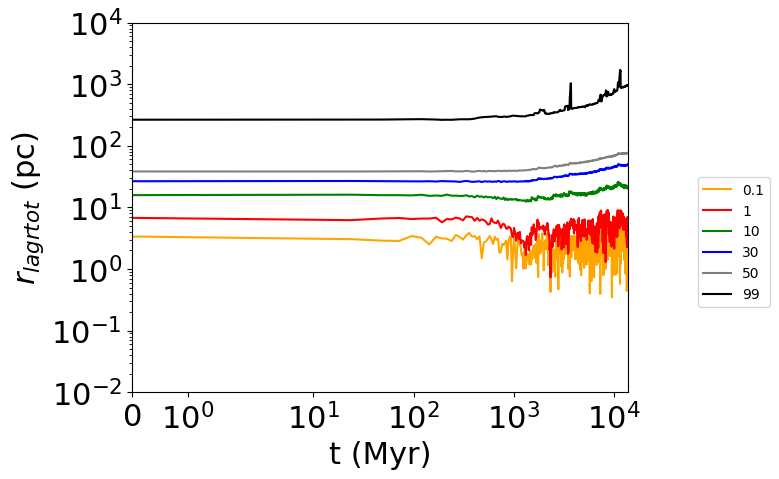}}
     {}
&
\subfloat{\includegraphics[width=60mm]{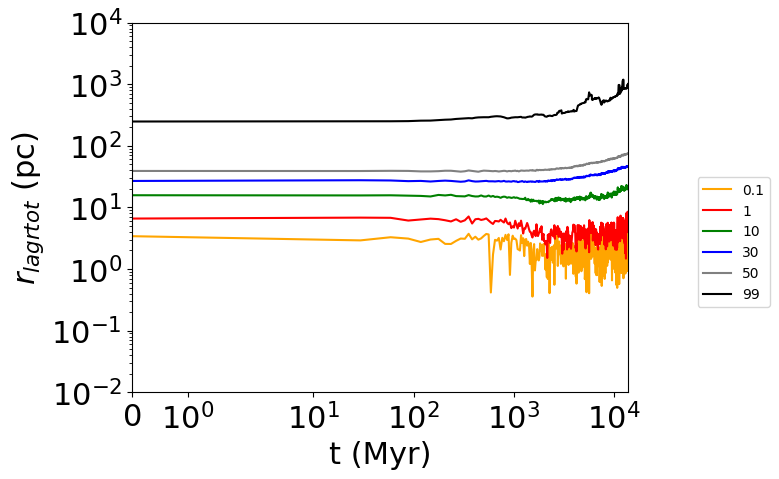}}
     {}
&
\subfloat{\includegraphics[width=60mm]{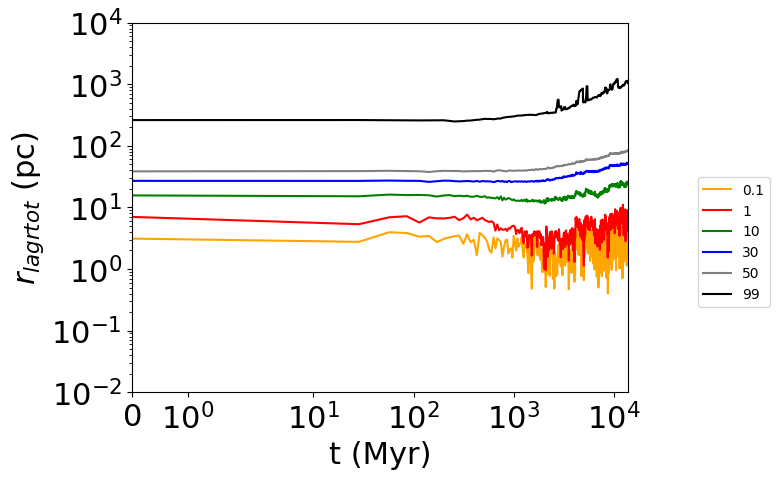}}
     {}
     \\
     \hline
\subfloat{\includegraphics[width=60mm]{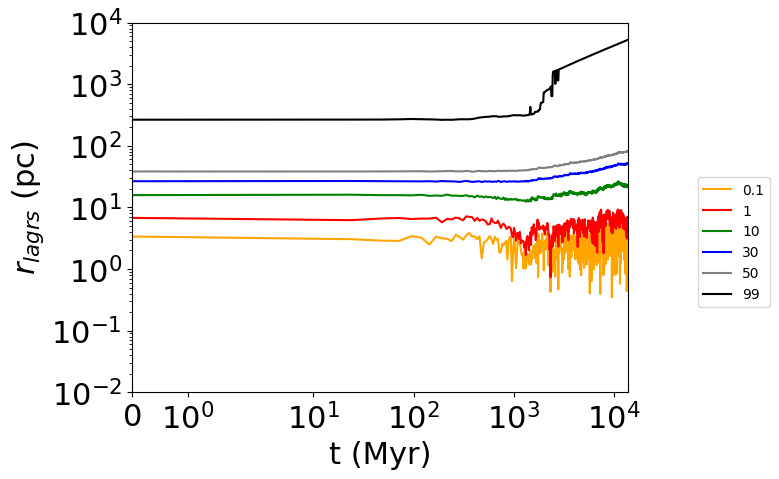}}
     {}
     &
     \subfloat{\includegraphics[width=60mm]{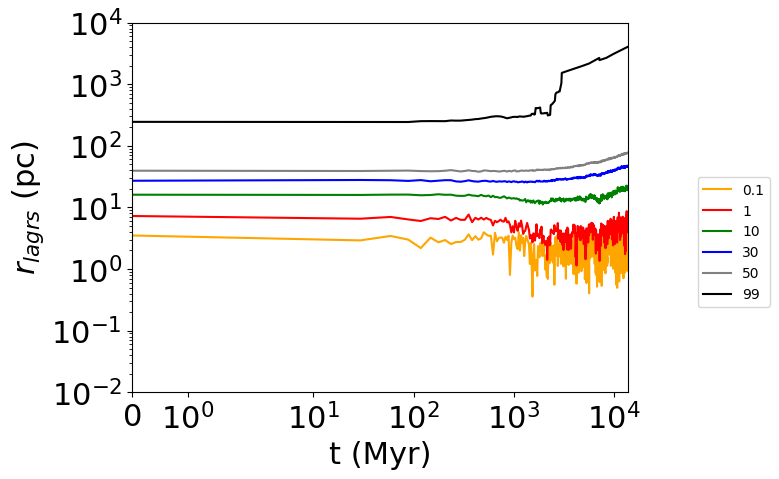}}
     {}
&
\subfloat{\includegraphics[width=60mm]{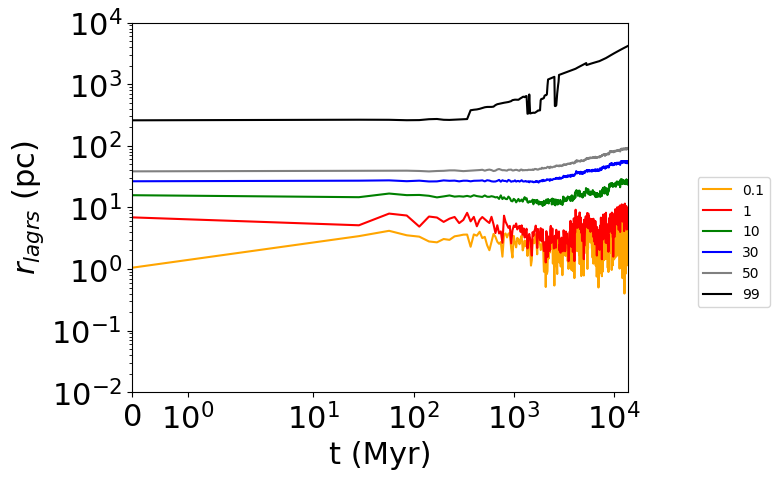}}
     {}
\\
\hline
\subfloat{\includegraphics[width=60mm]{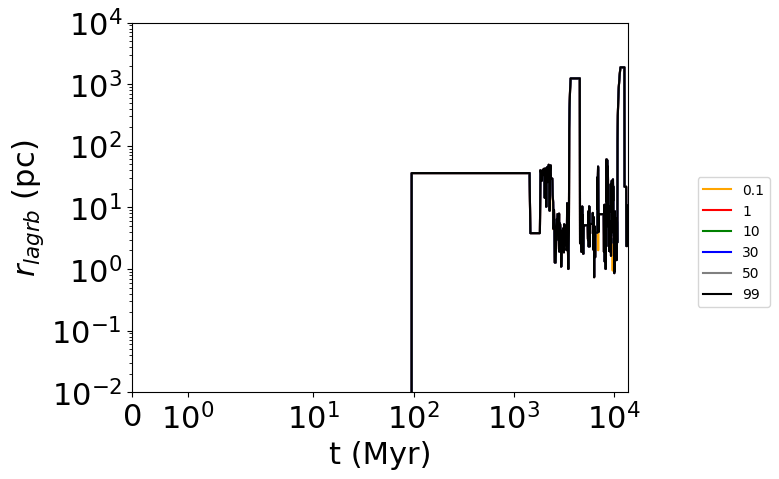}}
     {}
&
\subfloat{\includegraphics[width=60mm]{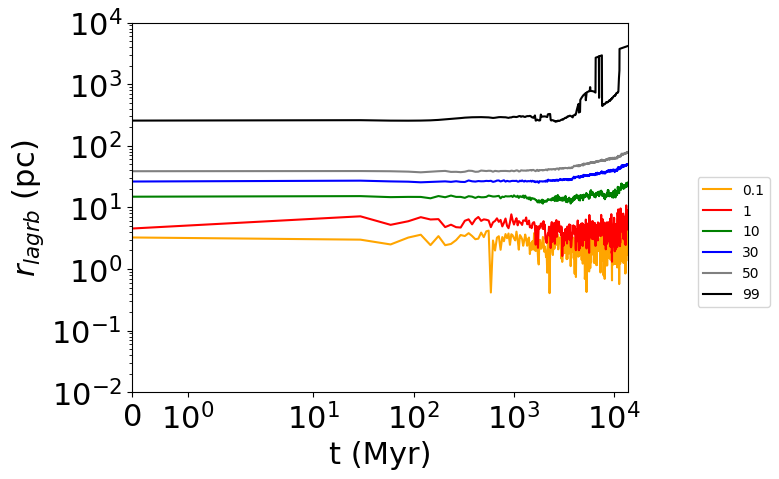}}
     {}
     &
\subfloat{\includegraphics[width=60mm]{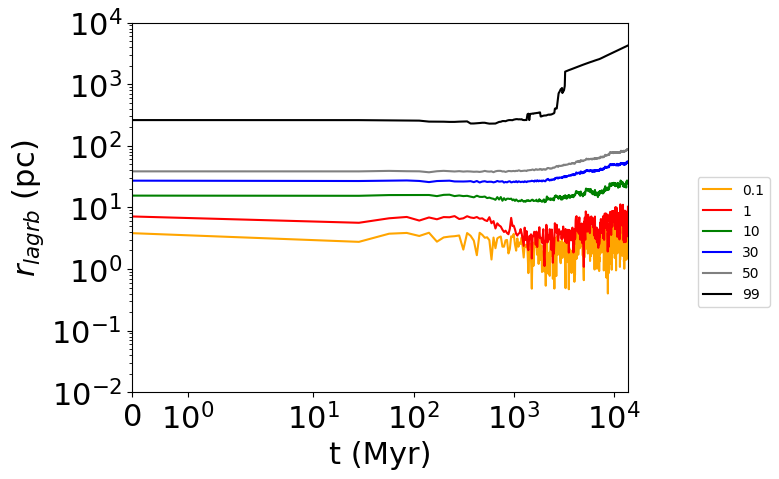}}
     {}
\\
\hline
\end{tabular}
}
\caption{Lagrangian radii for all cluster components (top row), single components (middle row), and binary components (bottom row) for models U0 (left column), U20 (middle), and U40  (right). In the bottom-left panel, we show only the 100\% Lagrangian radius, due to the presence of extremely few dynamical binaries. As expected, there are no major differences in the evolution of the components in the mass segregation and expansion of the UFD. }
\end{figure}
\clearpage

\section{HR diagrams of model U30}\label{sec:HRU30}
Figure B.1 presents HR diagrams of model U30 at different times, as labeled.
\begin{figure}[!ht]
\centering
\resizebox{\textwidth}{!}{\begin{tabular}{|c|c|c|}
\hline
\subfloat{\includegraphics[width=60mm]{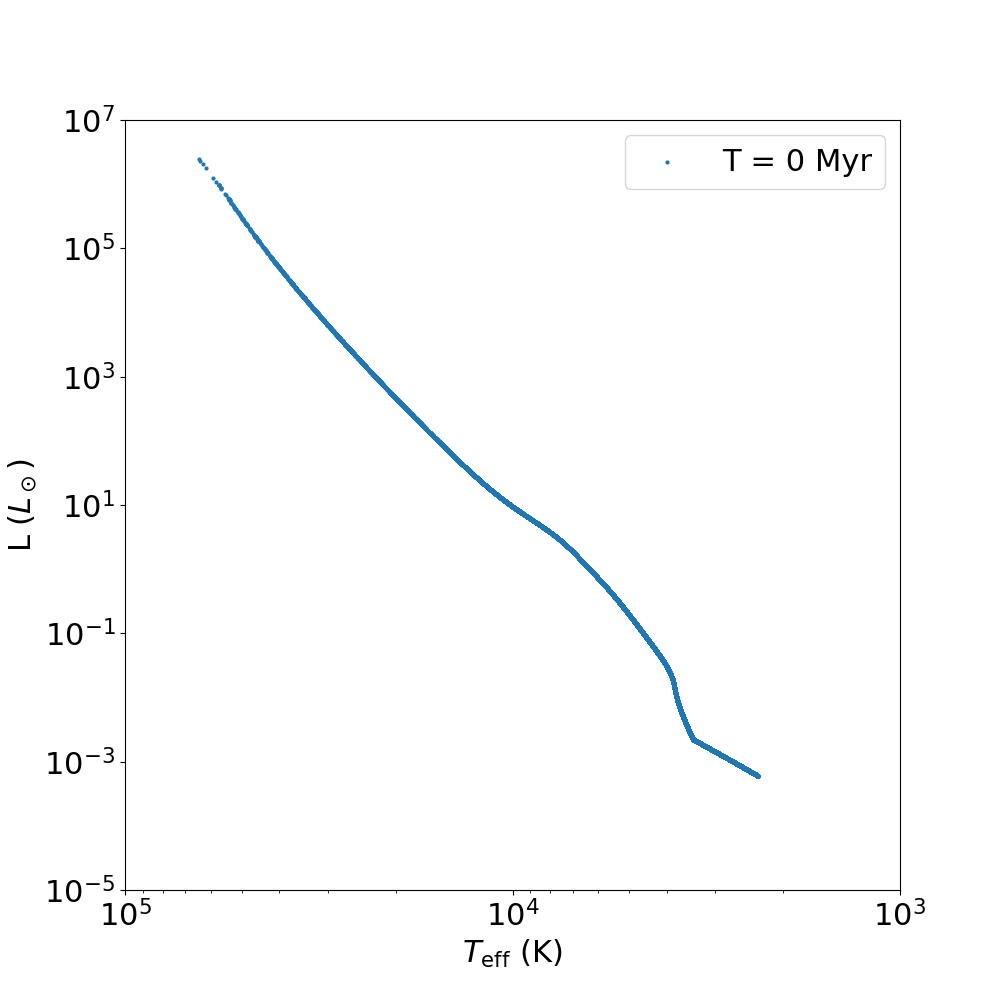}}
     {}
&
\subfloat{\includegraphics[width=60mm]{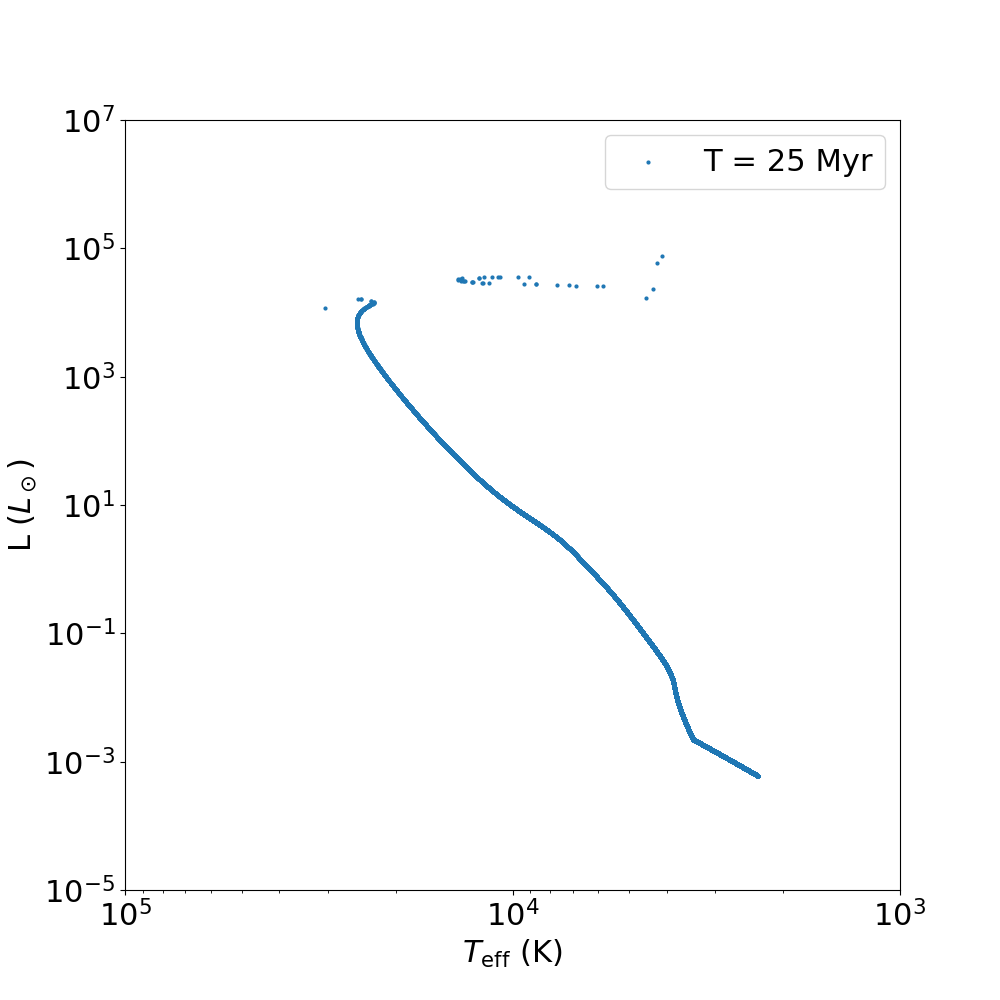}}
     {}
     &
     \subfloat{\includegraphics[width=60mm]{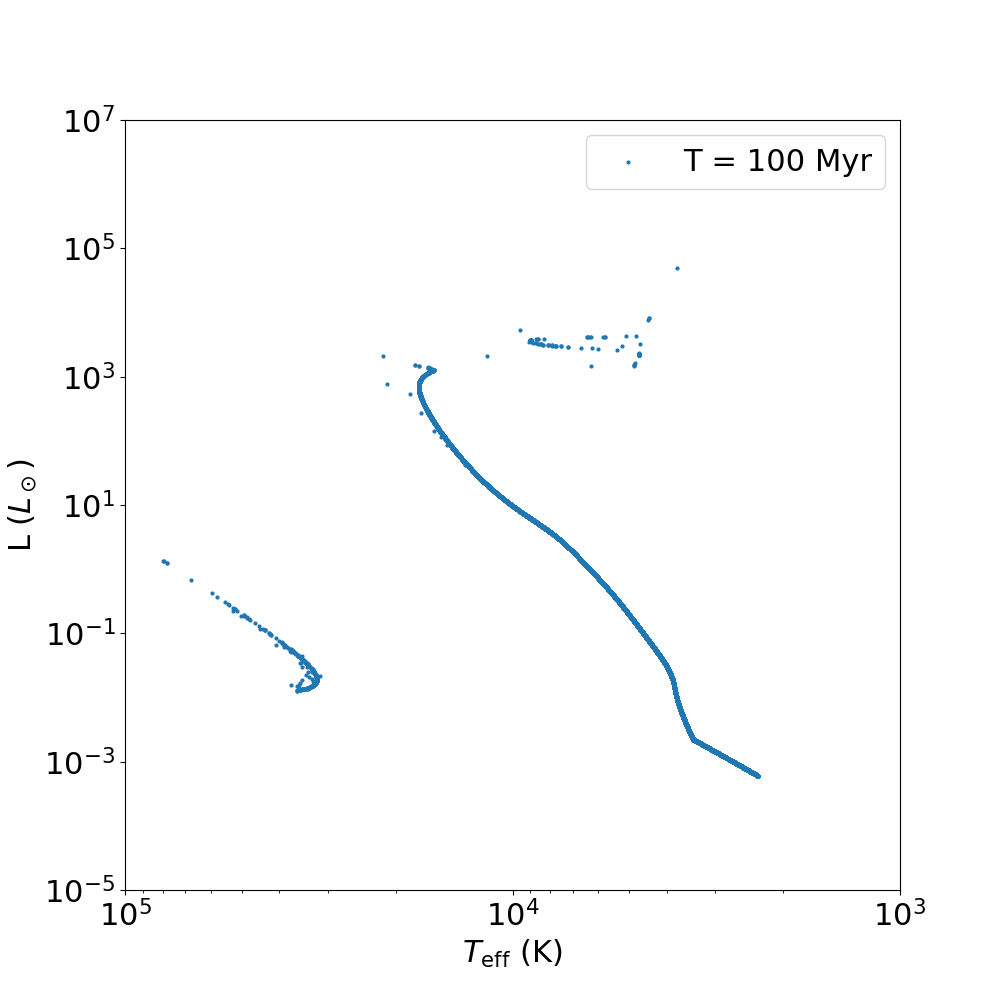}}
     {}
\\
\hline
\subfloat{\includegraphics[width=60mm]{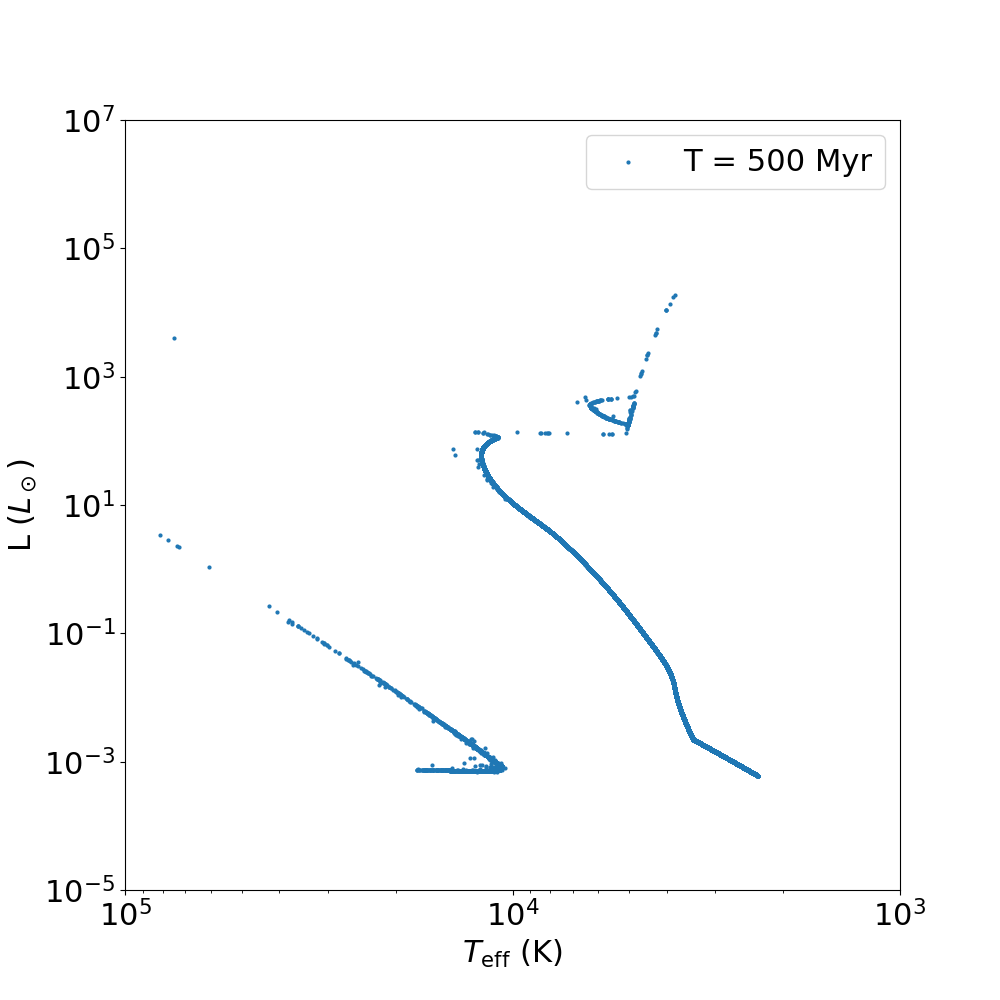}}
     {}
     &
     \subfloat{\includegraphics[width=60mm]{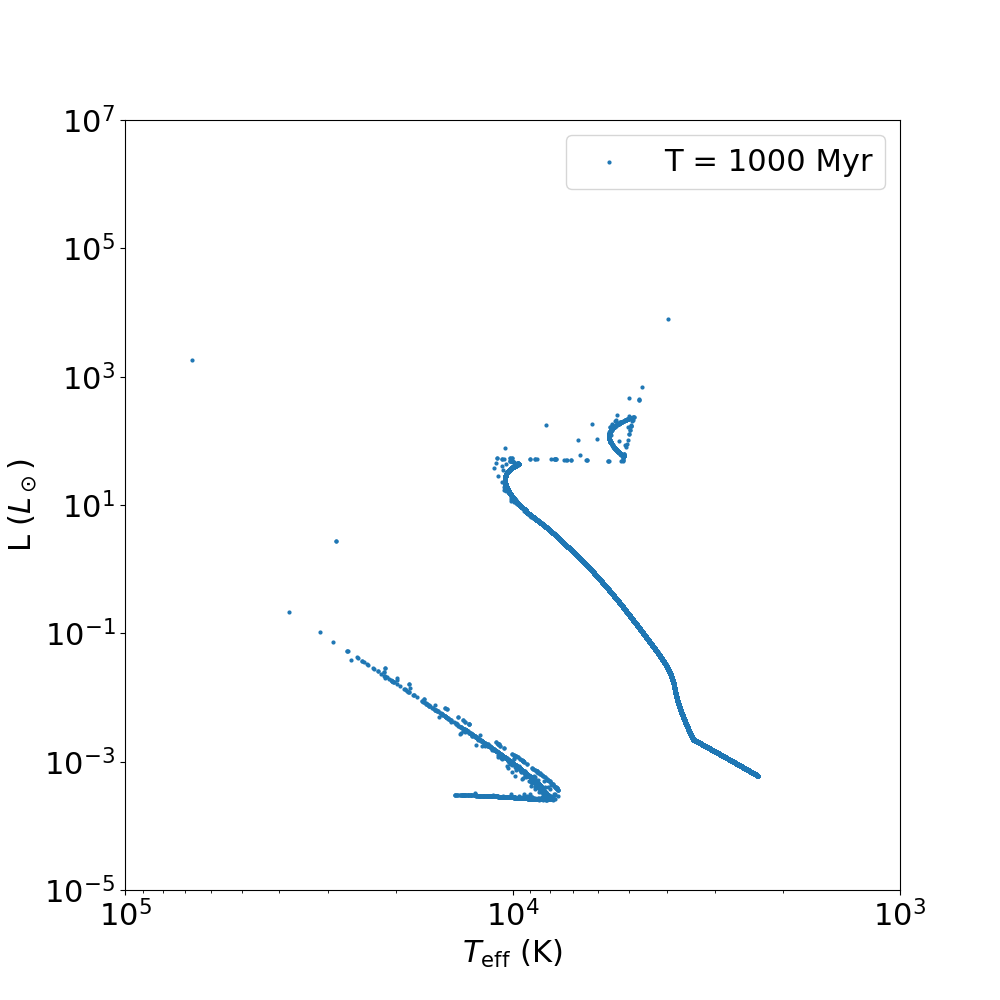}}
     {}
&
\subfloat{\includegraphics[width=60mm]{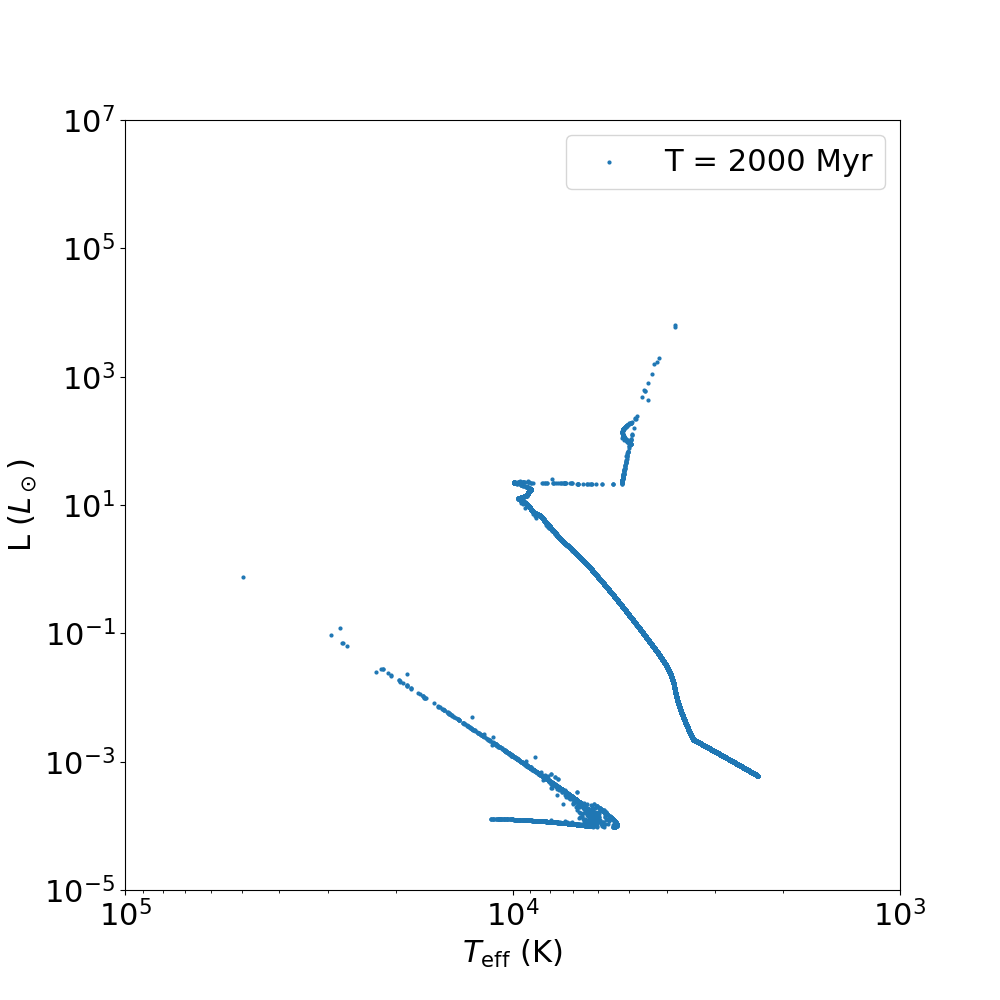}}
     {}
\\
\hline
\subfloat{\includegraphics[width=60mm]{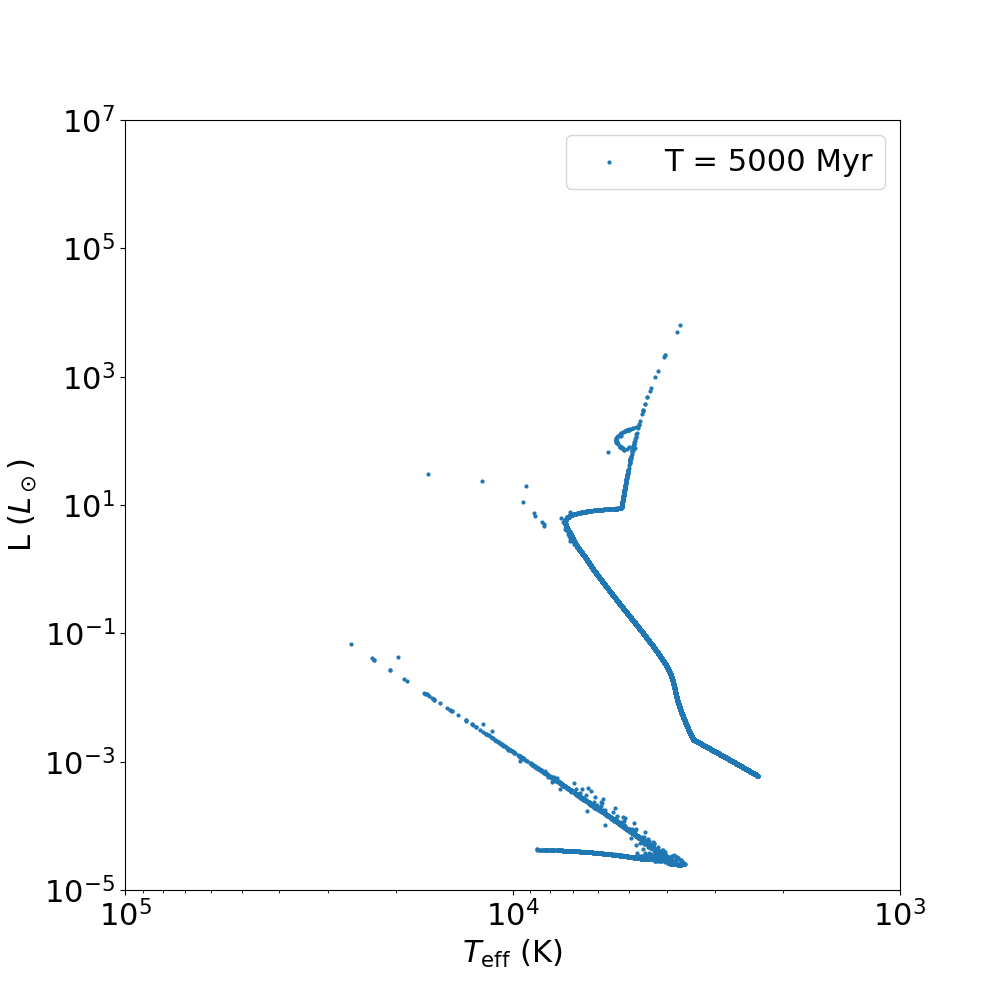}}
     {}
&
\subfloat{\includegraphics[width=60mm]{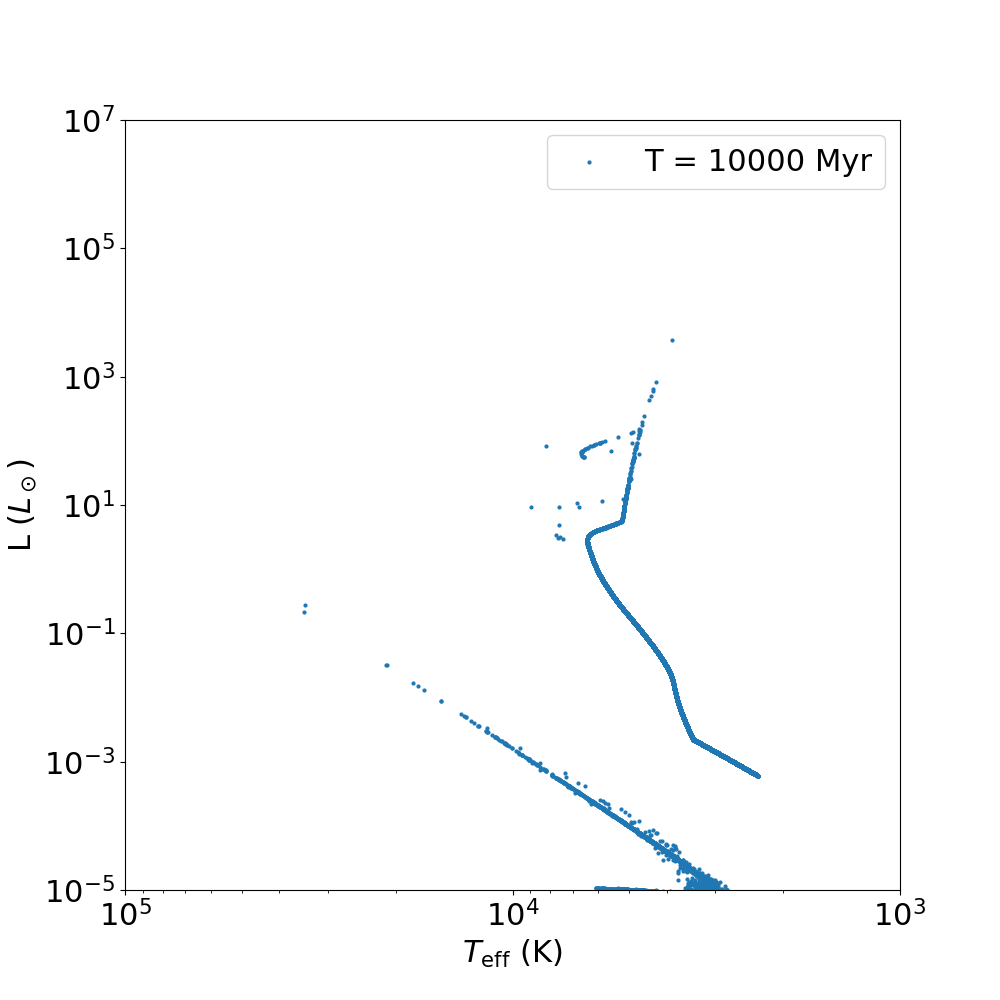}}
     {}
&
   \subfloat{\includegraphics[width=60mm]{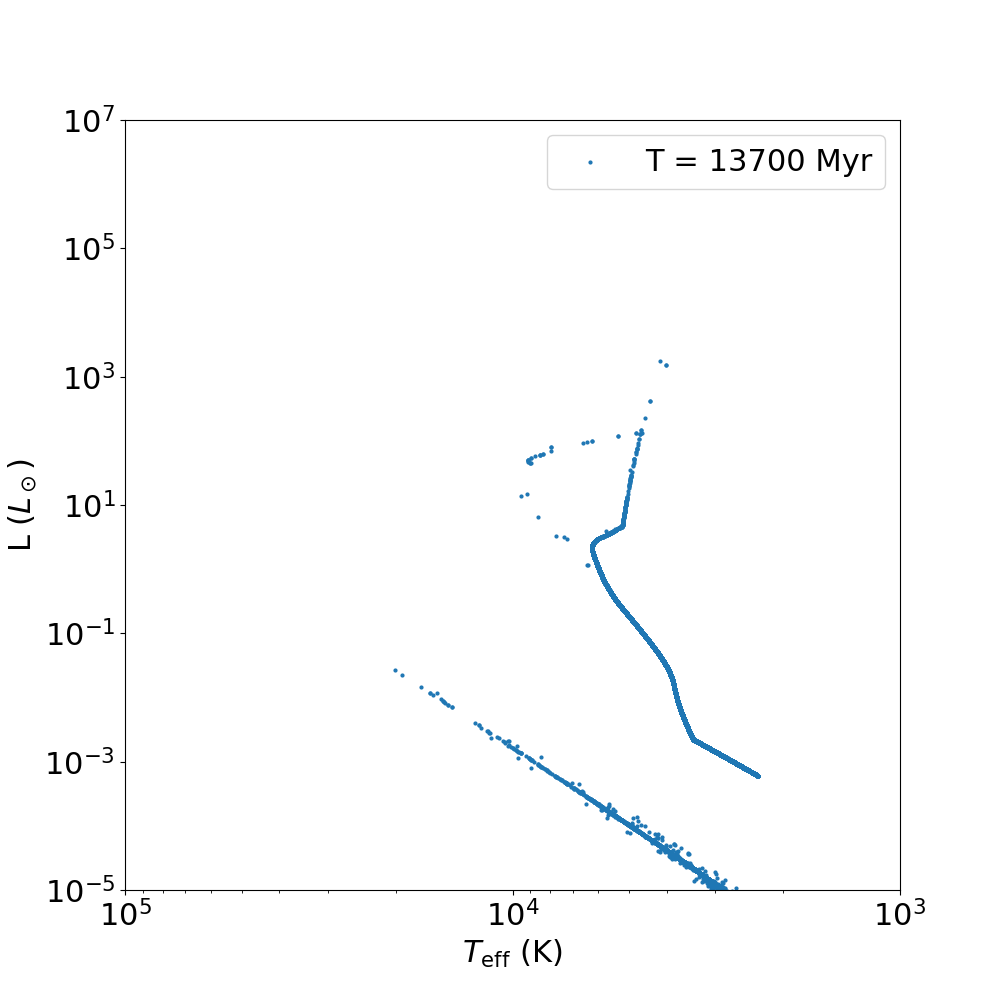}}
    {}
   \\
\hline
\end{tabular}}
\caption{HR diagrams for 0, 25, 100, 500, 1000, 2000, 5000, 10000, and 13700 Myr (as labeled) for model U30. }\label{HRtimes}
\end{figure}

\clearpage
\section{$\sigma_{\rm L}$ comparison between observations and our models}

In Fig. C.1 we compare the observations of  UFD present in Supplement Table 1 of \cite{sim19}, where  $M_{\rm V} > -7.7$ as a condition for identification as UFD is used. We include our six models, after properly deriving the integrated $V$ magnitude from our bolometric magnitudes. Our data are taken at the end of the simulation (Hubble time), except for model U50 (stopped at around 3\,000 Myr). Our UFD models are in the high luminosity tail of the plot. 
\newpage
   \begin{figure}[!ht]
   \centering
   \includegraphics[width=1\columnwidth]{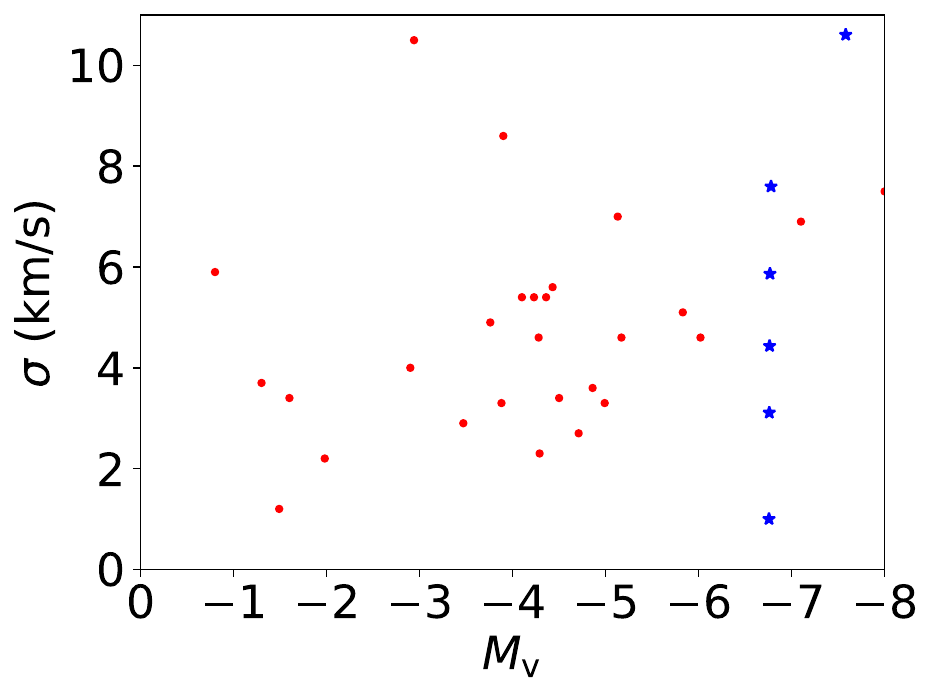}\\
      \caption{Velocity dispersion of UFDs as defined by \citet[red dots]{sim19} and our models (blue stars; U0 is the lowest and U50 the highest). Model U50 stopped at 3000 Myr.}
         \label{fig:sigmavsstars}
   \end{figure}

\end{appendix}

\end{document}